\documentclass[preprint,pre,aps,amsfonts,amsmath,showpacs,superscriptaddress,nofootinbib]{revtex4}
\usepackage{amsmath,amssymb,amscd}
\usepackage{graphicx}
\usepackage{bm}
\begin{document}
\title{Bose-Einstein Condensation of a Gaussian Random Field in the Thermodynamic Limit}
\author{Philippe Mounaix}
\email{mounaix@cpht.polytechnique.fr}
\affiliation{Centre de Physique Th\'eorique, UMR 7644 du CNRS, Ecole
Polytechnique, 91128 Palaiseau Cedex, France.}
\author{Satya N. Majumdar}
\email{majumdar@lptms.u-psud.fr}
\affiliation{Laboratoire de Physique Th\'eorique et Mod\`eles Statistiques, Universit\'e Paris-Sud, B\^at 100, 91405 Orsay Cedex, France.}
\author{Abhimanyu Banerjee}
\email{abhib@iitk.ac.in}
\affiliation{Department of Physics, Indian Institute of Technology Kanpur, Kanpur 208016, India.}
\date{\today}
\begin{abstract}
We derive the criterion for the Bose-Einstein condensation (BEC) of a Gaussian field $\varphi$ (real or complex) in the thermodynamic limit. The field is characterized by its covariance function and the control parameter is the intensity $u=\|\varphi\|_2^2/V$, where $V$ is the volume of the box containing the field. We show that for any dimension $d$ (including $d=1$), there is a class of covariance functions for which $\varphi$ exhibits a BEC as $u$ is increased through a critical value $u_c$. In this case, we investigate the probability distribution of the part of $u$ contained in the condensate. We show that depending on the parameters characterizing the covariance function and the dimension $d$, there can be two distinct types of condensate: a Gaussian distributed ``normal" condensate with fluctuations scaling as $1/\sqrt{V}$, and a non Gaussian distributed ``anomalous" condensate. A detailed analysis of the anomalous condensate is performed for a one-dimensional system ($d=1$). Extending this one-dimensional analysis to exactly the point of transition between normal and anomalous condensations, we find that the condensate at the transition point is still Gaussian distributed but with anomalously large fluctuations scaling as $\sqrt{\ln(L)/L}$, where $L$ is the system length. The conditional spectral density of $\varphi$, knowing $u$, is given for all the regimes (with and without BEC).
\end{abstract}
\maketitle
%
%
\section{Introduction}\label{sec1}
This paper is devoted to the concentration properties of a Gaussian random field $\varphi$ (real or complex) in the ``thermodynamic" limit, $V\rightarrow +\infty$ with fixed intensity $u=\|\varphi\|_2^2/V$, where $V$ is the volume of the box containing the field and $\|\varphi\|_2^2=\int_{\rm box}\vert\varphi(x)\vert^2\, d^d x$. For a finite system (with given $V<+\infty$), the question was first addressed in\ \cite{MD} heuristically and numerically in the context of laser-plasma interaction physics with spatially smoothed laser beams and, more generally, of linear amplification in systems driven by the square of a Gaussian noise. The results of\ \cite{MD} were extended and given a mathematically rigorous meaning in\ \cite{MC}.

Write $C_{\varphi\varphi^\ast}$ the covariance of $\varphi$ in a finite box with given $V<+\infty$, and $\kappa_1$ the largest eigenvalue of $C_{\varphi\varphi^\ast}$. The main result of\ \cite{MD,MC} states that $\varphi$ concentrates onto the eigenspace associated with $\kappa_1$ as $\|\varphi\|_2^2$ gets large. By revealing that the realizations of a Gaussian field that may cause the breakdown of a linear amplifier are {\it delocalized} modes, this result overturned conventional wisdom
\footnote{Most theoretical models dealing with stochastic amplification beyond the perturbative regime are hot spot models. In view of this result, the implicit assumption about the leading role of hot spots underlying all these models should be carefully reexamined.}
that breakdown is due to {\it localized} high-intensity peaks of the Gaussian field (the so-called high-intensity speckles, or hot spots\ \cite{RD}).

One of the goals of this paper is to study and provide a physical interpretation of this concentration property of a Gaussian field in the thermodynamic limit. We show that the emergence of a delocalized mode as the intensity $u$ increases beyond a critical value $u_c$ is similar to the Bose-Einstein condensation in an ideal Bose gas: the eigenspace associated with the largest eigenvalue $\kappa_1$ plays the role of the ground state in the Bose gas. Technically, this is also very similar to the spherical model of a ferromagnet. In that model, the nearest-neighbor ferromagnetic interaction of continuous spins $s_i\in{\mathbb R}$ on a lattice is supplemented with the global constraint $\sum_i s_i^2=N$, where $N$ is the number of lattice sites\ \cite{BK}. For large $N$ and in three or higher dimensions, the Fourier component of the spin field with $k=0$ gets thermodynamically populated as one reduces the  temperature below a critical temperature $T_c$, signalling the onset of a global nonzero magnetization (a delocalized mode). The mechanism of this phase transition is similar to the Bose-Einstein condensation\ \cite{GB}. In our problem, the role of the constraint $\sum_i s_i^2=N$ is played by $\|\varphi\|_2^2=\int_{\rm box}\vert\varphi(x)\vert^2\, d^d x=uV$, with fixed $u$. Our work thus makes a nice link between the concentration properties of random fields with fixed intensity (of much interest in e.g. laser-plasma physics) and the traditional Bose-Einstein condensation in statistical and atomic physics. Indeed, it can be shown that for $u$ large enough, the only contribution of the eigenspace associated with $\kappa_1$ is greater than the one of all the other eigenmodes of $C_{\varphi\varphi^\ast}$ (which remains bounded)\ \cite{MC}. Such a behavior is typical of a Bose-Einstein condensation onto the eigenspace associated with $\kappa_1$.

Of course, like any other phase transition, no sharp condensation can occur in a finite size system. The thermodynamic limit must be taken to get unambiguous results. In this limit, one is faced with the problem that more and more eigenvalues of $C_{\varphi\varphi^\ast}$ get closer and closer to $\kappa_1$ as $V\rightarrow +\infty$, which makes it difficult to tell them apart from $\kappa_1$ and may jeopardize condensation by leading to a concentration onto a larger space than the eigenspace associated with $\kappa_1$. For a homogeneous field\footnote{i.e. with correlation function $C(x,y)=C(x-y)$. Here and in the following, we use the word ``homogeneous" with the meaning of statistically invariant by translation in real or Fourier space, depending on the context (this will be specified explicitely in the text in case of ambiguity).}, one expects that issue to be all the more acute as the density of states at large wavelengths, close to the ground state (here, the condensate), is large. This will be the case at low space dimensionality $d$. Such a dimensional effect is well-known in traditional Bose-Einstein condensation of an ideal Bose gas which needs $d\ge 3$ to exist.

It is useful to summarize our main results. In this paper, we derive the criterion for the concentration of the Gaussian field $\varphi$ (real or complex) to turn into a Bose-Einstein condensation as the thermodynamic limit is taken. The Gaussian field is characterized by its covariance function and we have the intensity $u=\|\varphi\|_2^2/V$ as the control parameter. We show that in any dimension $d$ (including, in particular, $d=1$), there exists a class of covariance functions for which the system exhibits a Bose-Einstein condensation as one increases $u$ through a critical value $u_c$. In this case, we investigate the precise ``shape" of the condensate, i.e. the probability distribution of the part of $u$ contained in the condensate. We show that depending on the parameters characterizing the covariance function and the spatial dimension $d$, there can be two distinct types of condensate: the first one is Gaussian distributed with normal fluctuations scaling as $1/\sqrt{V}$ (``normal" condensate), and the second one is not Gaussian distributed (``anomalous" condensate). A schematic phase diagram is presented in Fig. \ref{fig1}. A detailed analysis of the structure of the anomalous condensate is performed for a one-dimensional system ($d=1$). Extending this one-dimensional analysis to exactly the point of transition between normal and anomalous condensations, we find that the condensate at the transition point is still Gaussian distributed but with anomalously large fluctuations scaling as $\sqrt{\ln(L)/L}$, where $L$ is the system length (in this sense it can be termed ``anomalous" too).

We will see later that when expressed in terms of the Fourier components of the field, our problem is structurally very similar to the one-dimensional mass transport model introduced in\ \cite{EMZ1}, in particular the condensation properties in this model\ \cite{MEZ,EMZ}. There are however a couple of important differences. In the mass transport model, condensation happens in real space\ \cite{MEZ,EMZ}, whereas in our Gaussian-field model, condensation happens in Fourier space. Secondly, unlike in the mass transport model where condensation in real space occurs {\it homogeneously}, i.e. the condensate can form at any point, here condensation is {\it heterogeneous} in Fourier space: for the particular class of fields we consider, it forms only at the $k=0$ mode. This is more in line with the traditional Bose-Einstein condensation (see e.g. the review\ \cite{EH} for several other examples of homogeneous vs. heterogeneous condensation). Nevertheless, much of our present analysis can be performed along the same line as in\ \cite{MEZ,EMZ} in all the regimes, with and without condensation. Indeed, ``normal" and ``anomalous" condensates are also found in the homogeneous mass transport model\ \cite{MEZ,EMZ}, although the precise nature of the condensate found here differs from the one in the homogeneous mass transport model. It is worth mentioning that condensation has also been studied for zero-range types of processes on imhomogeneous networks\ \cite{WBBJ}, which is also structurally somewhat similar to our model.

The outline of the paper is as follows. In Section\ \ref{sec2} we specify the class of $\varphi$ we consider and we give some necessary definitions. Section\ \ref{sec3} deals with the conditions for occurence of Bose-Einstein condensation. The conditional spectral density of $\varphi$, knowing $u=\|\varphi\|_2^2/V$, is given in Section\ \ref{sec4} for all the regimes (with and without condensation). Finally, the structure of the condensate is studied in Section\ \ref{sec5} for a one-dimensional system in the different condensation regimes and extension to higher space dimensionality is briefly discussed.
%
%
\section{Model and definitions}\label{sec2}
Let $\Lambda$ be a bounded subset of $\mathbb{R}^d$. In the following we take for $\Lambda$ a $d$-dimensional torus of length $L$ and volume $V=L^d$. The results in the thermodynamic limit are expected to be independent of the boundary conditions and imposing periodic boundary conditions by considering a torus makes the calculations simpler without loss of generality. Consider a homogeneous Gaussian random field $\varphi$ in $\Lambda$ with zero mean and correlation function $C_{\varphi\varphi^\ast}(x,y)\equiv C(x,y)=C(x-y)$, supplemented with $C_{\varphi\varphi}=0$ if $\varphi$ is complex and normalized such that $C(0)=1$. Since $\Lambda$ is a torus, $C(x)$ is also periodic with period $L$. For every $n\in\mathbb{Z}^d$, define
\begin{equation}\label{eq1.1}
C_L(n)=\int_\Lambda C(x){\rm e}^{-2i\pi n\cdot x/L}d^dx.
\end{equation}
As a correlation function, $C(x)$ is positive definite, hence $C_L(n)\ge 0$. Moreover, $C(x)$ is assumed to be such that (i) $C_L(n)<C_L(0)$ for every $n\ne 0$ and every $L$ large enough, (ii) for every $k\in\mathbb{R}^d$ the limit
\begin{equation}\label{eq1.2}
\tilde{C}(k)=\lim_{\stackrel{L,n\rightarrow +\infty}{2\pi n/L\rightarrow k}}C_L(n),
\end{equation}
exists and $\tilde{C}(k)<\tilde{C}(0)$ for every $k\ne 0$, and (iii) $\tilde{C}(k)=\tilde{C}(\vert k\vert)$ with $\tilde{C}(k)\sim\tilde{C}(0)\left(1-a\vert k\vert^\alpha\right)$ for small $k$ (with $a,\, \alpha >0$).

For every $n\in\mathbb{Z}^d$, we define the Fourier coefficients of $\varphi$ as
\begin{equation}\label{cor1.1}
\varphi_L(n)=\frac{1}{\sqrt{V}}\int_\Lambda \varphi(x){\rm e}^{-2i\pi n\cdot x/L}d^dx.
\end{equation}
If $\varphi$ is complex, the $\varphi_L(n)$ are independent and the joint probability density function (pdf) of their real and imaginary parts is given by
\begin{equation}\label{pdf1}
p\left\lbrack\lbrace {\rm Re}\, \varphi_L(n),\, {\rm Im}\, \varphi_L(n)\rbrace\right\rbrack =
\prod_{n\in{\mathbb Z}^d}\frac{1}{\pi C_L(n)}\, \exp\left\lbrack -\frac{\vert\varphi_L(n)\vert^2}{C_L(n)}\right\rbrack .
\end{equation}
The interested reader will find the derivation of Expressions\ (\ref{pdf1}) and\ (\ref{pdf2}) in Appendix\ \ref{app1}. If $\varphi$ is real, the $\varphi_L(n)$ are independent inside one given half of ${\mathbb Z}^d$ but {\it not} from one to the other complementary halves, as they are linked by Hermitian symmetry $\varphi_L(-n)=\varphi_L(n)^\ast$. It follows that a real $\varphi$ is determined by half as many degrees of freedom as a complex $\varphi$: its Fourier coefficients in only one half of ${\mathbb Z}^d$, with joint pdf
\begin{eqnarray}\label{pdf2}
p\left\lbrack\lbrace \varphi_L(0),\, {\rm Re}\, \varphi_L(n),\, {\rm Im}\, \varphi_L(n)\rbrace\right\rbrack &=&
\frac{1}{\sqrt{2\pi C_L(0)}}\, \exp\left\lbrack -\frac{\varphi_L(0)^2}{2C_L(0)}\right\rbrack \nonumber \\
&\times&\prod_{n\in{\mathbb Z}_{1/2}^d}\frac{1}{\pi C_L(n)}\, \exp\left\lbrack -\frac{\vert\varphi_L(n)\vert^2}{C_L(n)}\right\rbrack ,
\end{eqnarray}
where ${\mathbb Z}_{1/2}^d$ is a given half of ${\mathbb Z}^d$ excluding the point $n=0$. Note that $\varphi_L(0)$ is necessarily real by Hermitian symmetry. Now, it is possible to recast $\|\varphi\|_2^2$ as a sum of independent random variables, which will prove very useful in the following. By Parseval's theorem one has
\begin{equation}\label{cor1.2}
\|\varphi\|_2^2=\sum_{n\in{\mathbb Z}^d}\vert\varphi_L(n)\vert^2.
\end{equation}
If $\varphi$ is complex, its Fourier coefficients are independent and\ (\ref{cor1.2}) is already a sum of independent random variables. If $\varphi$ is real, one rewrites\ (\ref{cor1.2}) as
\begin{equation}\label{cor1.3}
\|\varphi\|_2^2=\varphi_L(0)^2
+2\sum_{n\in{\mathbb Z}_{1/2}^d}\lbrack{\rm Re}\, \varphi_L(n)\rbrack^2
+2\sum_{n\in{\mathbb Z}_{1/2}^d}\lbrack{\rm Im}\, \varphi_L(n)\rbrack^2,
\end{equation}
which is also a sum of independent random variables. Thus, one can always write
\begin{equation}\label{cor1.4}
\|\varphi\|_2^2=\sum_{n\in{\mathbb Z}^d}m_n,
\end{equation}
where the $m_n$ are independent random variables given by quadratic forms of ${\rm Re}\, \varphi_L(n)$ and ${\rm Im}\, \varphi_L(n)$ according to either\ (\ref{cor1.2}) (complex $\varphi$) or\ (\ref{cor1.3}) (real $\varphi$). Namely, in the former case $m_n= \vert\varphi_L(n)\vert^2$, whereas in the latter case one takes $m_0=\varphi_L(0)^2$, $m_n=2\lbrack{\rm Re}\, \varphi_L(n)\rbrack^2$, and $m_{-n}=2\lbrack{\rm Im}\, \varphi_L(n)\rbrack^2$ for $n\in{\mathbb Z}_{1/2}^d$. Since the $m_n$ are independent, their joint pdf is a product measure straightforwardly obtained from either\ (\ref{pdf1}) (complex $\varphi$) or\ (\ref{pdf2}) (real $\varphi$). One finds
\begin{equation}\label{cor1.5}
p\left\lbrack\lbrace m_n\rbrace\right\rbrack =
\prod_{n\in{\mathbb Z}^d}p_n(m_n),
\end{equation}
with
\begin{equation}\label{cor1.5b}
p_n(m_n)=\frac{{\bm 1}_{\lbrace m_n>0\rbrace}}{\pi}\, 
\left(\frac{\pi}{\varepsilon C_L(n)}\right)^{1/\varepsilon}\, \frac{1}{m_n^{1-1/\varepsilon}}\, \exp\left\lbrack -\frac{m_n}{\varepsilon C_L(n)}\right\rbrack ,
\end{equation}
where $\varepsilon =1$ (resp. $\varepsilon =2$) when $\varphi$ is complex (resp. real)\footnote{A word of caution for the reader: the notation $\varepsilon$ should not be confused with the one of a small number.}. Replacing then $\|\varphi\|_2^2$ with\ (\ref{cor1.4}) in the expression $p(U)=\langle\delta(\|\varphi\|_2^2 -U)\rangle$ of the probability density of $\|\varphi\|_2^2$, where the average $\langle\cdot\rangle$ is over the product measure\ (\ref{cor1.5}), one obtains
\begin{equation}\label{cor1.6}
p(U)=\int_{\lbrace m_n\ge 0\rbrace}
\delta\left(U-\sum_{i\in{\mathbb Z}^d}m_i\right)\, 
\prod_{n\in{\mathbb Z}^d} p_n(m_n)\, dm_n.
\end{equation}

At this point, it may be interesting to shortly digress and point out in more detail the similarities and differences between our model and the mass transport model studied in Refs.\ \cite{EMZ1,MEZ,EMZ}. In the latter, there is positive mass variable $m_n$ at each site $n\in{\mathbb Z}$ of a one-dimensional lattice with periodic boundary conditions. The microscopic dynamics consists in transferring (or chipping) a random portion of mass from site $n$ to site $(n+1)$. The amount of mass, ${\tilde m}$, to be transferred from a site with mass $m$ to its neighbor is chosen from a prescribed distribution $K\left({\tilde m}\vert m\right)$ which is called the chipping kernel. The chipping kernel is homogeneous, i.e. does not depend on the site. The dynamics conserves the total mass $M=\sum_n m_n$. For a class of chipping kernels\ \cite{EMZ1}, the system reaches a stationary state in the long time limit where the joint pdf of masses becomes time independent with the simple form\ \cite{EMZ1}
\begin{equation}\label{cor1.7}
p\left\lbrack\lbrace m_n\rbrace\right\rbrack = \frac{1}{Z_N(M)}\, 
\delta\left(M-\sum_i m_i\right)\, \prod_n f(m_n),
\end{equation}
where
\begin{equation}\label{cor1.8}
Z_N(M)= \int_{\lbrace m_n\ge 0\rbrace}
\delta\left(M-\sum_i m_i\right)\, \prod_n f(m_n)\, dm_n,
\end{equation}
is the normalizing partition function, and $N$ is the number of lattice sites. The weight function $f(m)$, the same for each site $n$, is non-negative and depends on the chipping kernel $K\left({\tilde m}\vert m\right)$. By choosing this kernel appropriately, one can generate a whole class of weight functions $f(m)$. The choice $f(m)\sim m^{-\gamma}$ for large $m$ with $\gamma>2$ leads to condensation whereby a finite fraction of the total mass $M$ condenses onto a single site of the lattice when the density $\rho=M/N$ exceeds a critical value $\rho_c$\ \cite{MEZ,EMZ}.

The product measure structure in the two problems is similar and $p(U)$ given by\ (\ref{cor1.6}) is just the analogue of the partition function $Z_N(M)$ given by\ (\ref{cor1.8}). However, there is an important difference. In our case, the counterpart of the weight function in\ (\ref{cor1.6}), $p_n(m_n)$, depends explicitly on $n$ (through its dependence on the covariance function $C_L(n)$), whereas in the mass transport model the weight function $f(m_n)$ is identical for all sites $n$\ \cite{EMZ1,EMZ,MEZ}. In this sense, our model is heterogeneous in contrast to the homogeneous mass transport model.

After this digression, we return to the expression\ (\ref{cor1.6}) of $p(U)$ and express it in a more tractable form. The $m_n$-integrals in\ (\ref{cor1.6}) are most conveniently performed by using the Laplace representation of the delta function or, equivalently, by writing $p(U)$ as the Bromwich integral of its Laplace transform $\tilde{p}(\lambda)$ with
\begin{eqnarray}\label{eq1.3}
\tilde{p}(\lambda)&=&\int_0^{+\infty}p(U)\exp(-\lambda U)\, dU \nonumber \\
&=&\left\langle\exp\left(-\lambda\sum_{n\in{\mathbb Z}^d}m_n\right)\right\rangle \\
&=&\prod_{n\in{\mathbb Z}^d}\frac{1}{\lbrack 1+\varepsilon\lambda C_L(n)\rbrack^{1/\varepsilon}}, \nonumber
\end{eqnarray}
where we have used\ (\ref{cor1.4}) and performed the average over the product measure\ (\ref{cor1.5}), [recall that $\varepsilon =1$ (resp. $\varepsilon =2$) when $\varphi$ is complex (resp. real)]. Then, by inverse Laplace transform,
\begin{equation}\label{eq1.4}
p(U)=\int_{\mathcal L}{\rm e}^{\lambda U-\frac{1}{\varepsilon}\sum_{n\in{\mathbb Z}^d}\ln\lbrack 1+\varepsilon\lambda C_L(n)\rbrack}
\frac{d\lambda}{2i\pi},
\end{equation}
the integral being evaluated along a vertical line in the complex plane upward and to the right of all the singularities of the integrand. This contour of integration (or any of its continuous deformation within the analyticity domain of the integrand) defines a Bromwich contour generically denoted by ${\mathcal L}$. Making the shift $s=\lambda+1/\varepsilon C_L(0)$, one gets
\begin{equation}\label{eq1.5}
p(U)={\rm e}^{-\frac{U}{\varepsilon C_L(0)}}
\int_{\mathcal L}{\rm e}^{sU-\frac{1}{\varepsilon}
\sum_{n\in{\mathbb Z}^d}\ln\lbrack 1-C_L(n)/C_L(0)+\varepsilon sC_L(n)\rbrack}
\frac{ds}{2i\pi}.
\end{equation}
In the thermodynamic limit, it is more convenient to work with the intensive variable $u=\|\varphi\|_2^2/V$ instead of the extensive one $U=\|\varphi\|_2^2$. Defining
\begin{equation}\label{eq1.6}
S(s,u,V)=su-\frac{1}{\varepsilon V}
\sum_{n\in{\mathbb Z}^d}\ln\left\lbrack 1-\frac{C_L(n)}{C_L(0)}+\varepsilon sC_L(n)\right\rbrack ,
\end{equation}
and using\ (\ref{eq1.5}), one finds that the probability density of $u$ is given by the integral representation,
\begin{equation}\label{eq1.7}
p(u)=Vp(U=uV)=V{\rm e}^{-\frac{Vu}{\varepsilon C_L(0)}}
\int_{\mathcal L}{\rm e}^{VS(s,u,V)}\frac{ds}{2i\pi}.
\end{equation}
The leading asymptotic behavior of $p(u)$ in the large $V$ limit can then be obtained from a steepest descent analysis of\ (\ref{eq1.7}) in which $S(s,u,V)$ is approximated by
\begin{equation}\label{eq1.8}
S(s,u,V)\sim\left\lbrace
\begin{array}{lr}
su-I(s)&(V\rightarrow +\infty ,\ s\, {\rm fixed}),\\
su-I(s)-\frac{1}{\varepsilon V}\ln\left\lbrack\varepsilon s\tilde{C}(0)f(V)\right\rbrack&
(s\rightarrow 0,\, {\rm then}\, V\rightarrow +\infty),
\end{array}\right.
\end{equation}
where
\begin{equation}\label{cor1.9}
I(s)=\frac{1}{\varepsilon}\int
\ln\left\lbrack 1-\frac{\tilde{C}(k)}{\tilde{C}(0)}+\varepsilon s\tilde{C}(k)\right\rbrack\, \frac{d^d k}{(2\pi)^d},
\end{equation}
and $f(V)>0$, in the second line of Eq.\ (\ref{eq1.8}), is a non trivial correction resulting from the discreteness of the infrared modes around $k=0$. It turns out that in the sequence of limits $s\rightarrow 0$ then $V\rightarrow +\infty$, these modes cannot be described properly as a continuum, and a careful analysis of the sum in\ (\ref{eq1.6}) is required, leading to the correction $f(V)$. In one dimension, $f(L)$ can be obtained explicitely from the Euler-Maclaurin formula at lowest order applied to\ (\ref{eq1.6}). One finds $f(L)\sim L^\alpha$ with $\alpha =\lim_{k\rightarrow 0}\ln\vert\tilde{C}(k)-\tilde{C}(0)\vert /\ln\vert k\vert$. For higher dimensions, algebraic growth of $f(V)$ is also expected as $V\nearrow +\infty$. The specific expression of $f(V)$ will not be needed in the following.

It is convenient to express the function $I(s)$ in\ (\ref{cor1.9}) in the following form
\begin{equation}\label{eq1.9}
I(s)=I(0)+\frac{1}{\varepsilon}\int
\ln\left( 1+\frac{s}{g(k)}\right)\, \frac{d^d k}{(2\pi)^d},
\end{equation}
with
\begin{equation}\label{eq1.10}
g(k)=\frac{1}{\varepsilon}\left\lbrack\frac{1}{\tilde{C}(k)}-\frac{1}{\tilde{C}(0)}\right\rbrack .
\end{equation}
For later purposes we also define a critical intensity, $u_c\le +\infty$, by
\begin{equation}\label{eq1.11}
u_c=I^\prime(0)=\frac{1}{\varepsilon}\int\frac{1}{g(k)}\, \frac{d^d k}{(2\pi)^d}.
\end{equation}
The existence of the integral on the right-hand side of\ (\ref{eq1.11}) depends on the small $k$ behavior of $g(k)$. For the class of $\varphi$ we consider, one has
\begin{equation}\label{eq1.12}
g(k)\sim\frac{a\vert k\vert^\alpha}{\varepsilon\tilde{C}(0)}\ \ \ \ (k\rightarrow 0),
\end{equation}
from which it follows that $u_c <+\infty$ (resp. $u_c =+\infty$) if $\alpha <d$ (resp. $\alpha \ge d$).

The expression of $p(u)$ depends on which limit in Eq.\ (\ref{eq1.8}) must be used in Eq.\ (\ref{eq1.7}). The right choice is determined by the contribution of the $k=0$ mode to the sum\ (\ref{eq1.6}) in the thermodynamic limit : if $u<u_c$, it is negligible and the proper asymptotic expression for $S(s,u,V)$ is the first line of\ (\ref{eq1.8}) ; if $u>u_c$, it becomes significant and the proper expression for $S(s,u,V)$ is the second line of\ (\ref{eq1.8}). (See the beginning of Secs.\ \ref{sec3.1} and\ \ref{sec3.5}, the discussion in Sec.\ \ref{sec5.2}, and Appendix\ \ref{app2} for further details). It is possible to write the corresponding two different expressions of $p(u)$ into a single concise formula by introducing a ``tag", $\sigma =0$ or $1$, indicating which limit in Eq.\ (\ref{eq1.8}) is considered. More precisely, injecting either line of\ (\ref{eq1.8}) into\ (\ref{eq1.7}) one gets
\begin{equation}\label{eq1.13}
p(u)\sim V\left\lbrack\varepsilon\tilde{C}(0)f(V)\right\rbrack^{-\sigma /\varepsilon}
{\rm e}^{-\frac{Vu}{\varepsilon\tilde{C}(0)}}
\int_{\mathcal L}s^{-\sigma /\varepsilon}{\rm e}^{V\lbrack su-I(s)\rbrack}\frac{ds}{2i\pi},
\end{equation}
where taking $\sigma =0$ (resp. $\sigma =1$) gives the expression of $p(u)$ obtained from to the first (resp. second) line of\ (\ref{eq1.8}). The criterion for a condensation of $\varphi$ can then be found by analyzing the function
\begin{equation}\label{eq1.14}
\Xi(u)=-\lim_{V\rightarrow +\infty}\frac{1}{V}\ln p(u).
\end{equation}
A condensation transition of $\varphi$ at some $u=u_{cond}$ is indicated by the fact that $\Xi(u)$ is not analytic at $u=u_{cond}$. Replacing $p(u)$ with\ (\ref{eq1.13}), one gets
\begin{equation}\label{eq1.15}
\Xi(u)=\frac{u}{\varepsilon\tilde{C}(0)}-\lim_{V\rightarrow +\infty}\frac{1}{V}\ln \int_{\mathcal L}s^{-\sigma /\varepsilon}{\rm e}^{V\lbrack su-I(s)\rbrack}\frac{ds}{2i\pi},
\end{equation}
which is the expression of $\Xi(u)$ we will use in the next section.
%
%
\section{Conditions for Bose-Einstein condensation}\label{sec3}
\subsection{$\bm{u<u_c}$}\label{sec3.1}
For fixed $u<u_c$ there always is one saddle point $s(u)>0$, solution to $u=I^\prime (s(u))$, which determines the leading exponential behavior of the Bromwich integral on the right-hand side of\ (\ref{eq1.15}). In this case $\Xi(u)$ reduces to
\begin{equation}\label{eq2.1}
\Xi(u)=\frac{u}{\varepsilon\tilde{C}(0)}+I(s(u))-us(u).
\end{equation}
Note that $s(u)$ being fixed, $S(s,u,V)$ is given by the first Eq.\ (\ref{eq1.8}), which corresponds to $\sigma =0$ in Eqs.\ (\ref{eq1.13}) and\ (\ref{eq1.15}).

By analyticity of $I(s)$ in ${\mathbb C}^+=\lbrace s\in {\mathbb C}:{\rm Re}\, s>0\rbrace$ and the inverse function theorem for complex functions, $s(u)$ is analytic in the interval $0<u<u_c$. It follows that $\Xi(u)$ is also analytic in the same interval and no condensation transition of $\varphi$ is to be expected for $0<u<u_c$. Since  $u_c =+\infty$ if $\alpha \ge d$, one can already conclude that Bose-Einstein condensation of $\varphi$ needs $\alpha <d$ to exist.

We now take a closer look at the behavior of $\Xi(u)$ for $\alpha <d$ and $u$ close to $u_c<+\infty$. There are three different cases to consider: $\alpha <d/2$, $d/2<\alpha <d$, and $\alpha =d/2$.
%
%
\subsubsection{$\alpha <d/2$ and $u\rightarrow u_c^{-}$}\label{sec3.2}
Write $\Delta(u)^2=-I^{\prime\prime}(s(u))>0$. From the small $k$ behavior of $g(k)$, Eq.\ (\ref{eq1.12}), it can be checked that $\Delta (0) <+\infty$ if $\alpha <d/2$. When $u$ gets close to $u_c$ from below, $s(u)$ gets small and the saddle point equation can be expanded as
\begin{eqnarray*}
u=I^\prime (s(u))&\simeq&I^\prime (0)+I^{\prime\prime}(0)s(u) \\
&=&u_c-\Delta(0)^2 s(u),
\end{eqnarray*}
yielding
\begin{equation*}
s(u)\simeq\frac{u_c-u}{\Delta(0)^2},
\end{equation*}
and, to the same accuracy,
\begin{eqnarray}\label{eq2.2}
I(s(u))-us(u)&\simeq&\left\lbrack I(0)+I^\prime (0)s(u)+\frac{1}{2}I^{\prime\prime}(0)s(u)^2\right\rbrack
-\left\lbrack u_cs(u)+(u-u_c)s(u)\right\rbrack \nonumber \\
&=&I(0)+(u_c-u)s(u)-\frac{1}{2}\Delta(0)^2s(u)^2 \nonumber \\
&\simeq& I(0)+\frac{(u_c-u)^2}{2\Delta(0)^2}.
\end{eqnarray}
Injecting\ (\ref{eq2.2}) into the right-hand side of\ (\ref{eq2.1}), one gets
\begin{equation}\label{eq2.3}
\Xi(u)=I(0)+\frac{u}{\varepsilon\tilde{C}(0)}+\frac{(u_c-u)^2}{2\Delta(0)^2}.
\end{equation}
%
%
\subsubsection{$d/2<\alpha <d$ and $u\rightarrow u_c^{-}$}\label{sec3.3}
In this case $\Delta(0)=+\infty$ (because $g(k)^{-2}$ is not infrared integrable) and a more careful analysis is needed. Take $s>0$ real and write\ (\ref{eq1.9}) as
\begin{eqnarray}\label{eq2.4}
I(s)&=&
I(0)+u_c s
+\frac{1}{\varepsilon}\int\left\lbrack\ln\left(1+\frac{s}{g(k)}\right)-\frac{s}{g(k)}\right\rbrack\, \frac{d^d k}{(2\pi)^d} \\
&=&I(0)+u_c s
+\frac{s^{d/\alpha}}{\varepsilon}\int\left\lbrack\ln\left(1+\frac{s}{g(s^{1/\alpha}q)}\right)-\frac{s}{g(s^{1/\alpha}q)}\right\rbrack\, \frac{d^d q}{(2\pi)^d}, \nonumber
\end{eqnarray}
where we have made the change of variable $k=s^{1/\alpha}q$. Writing the last term at the lowest order in $s$, one gets
\begin{equation}\label{eq2.5}
I(s)=I(0)+u_c s-J s^{d/\alpha}+o(\vert s\vert^{d/\alpha}),
\end{equation}
with
\begin{equation}\label{eq2.6}
J=\frac{1}{\varepsilon}\int\left\lbrack
\frac{\varepsilon\tilde{C}(0)}{a\vert q\vert^\alpha}-
\ln\left(1+\frac{\varepsilon\tilde{C}(0)}{a\vert q\vert^\alpha}\right)\right\rbrack\, \frac{d^d q}{(2\pi)^d}.
\end{equation}
Note that the inequalities $d/2<\alpha <d$ ensure $J$ to be both infrared and ultraviolet convergent. Since $I(s)$ is analytic in ${\mathbb C}^+$, the expansion\ (\ref{eq2.5}) holds not only for small $s$ in ${\mathbb R}^+$ but also for small $s$ in ${\mathbb C}^+$.

One can then follow the same line as in Sec.\ \ref{sec3.2} with $I(s)$ now given by\ (\ref{eq2.5}). For $u<u_c$ close to $u_c$, the saddle point is solution to
\begin{equation*}
u=I^\prime (s(u))\simeq u_c-\frac{dJ}{\alpha}s(u)^{d/\alpha -1},
\end{equation*}
which gives
\begin{equation*}
s(u)=\left\lbrack\frac{\alpha (u_c-u)}{dJ}\right\rbrack^{\frac{\alpha}{d-\alpha}},
\end{equation*}
and, to the same accuracy,
\begin{eqnarray}\label{eq2.7}
I(s(u))-us(u)&\simeq&\left\lbrack I(0)+u_c s(u)-J s(u)^{d/\alpha}\right\rbrack
-\left\lbrack u_cs(u)+(u-u_c)s(u)\right\rbrack \nonumber \\
&=&I(0)+(u_c-u)s(u)-J s(u)^{d/\alpha} \nonumber \\
&\simeq& I(0)+\left(1-\frac{\alpha}{d}\right)\, \left(\frac{\alpha}{dJ}\right)^{\frac{\alpha}{d-\alpha}} (u_c-u)^{\frac{d}{d-\alpha}}.
\end{eqnarray}
Injecting\ (\ref{eq2.7}) into the right-hand side of\ (\ref{eq2.1}), one gets
\begin{equation}\label{eq2.8}
\Xi(u)=I(0)+\frac{u}{\varepsilon\tilde{C}(0)}+\left(1-\frac{\alpha}{d}\right)\, \left(\frac{\alpha}{dJ}\right)^{\frac{\alpha}{d-\alpha}}
(u_c-u)^{\frac{d}{d-\alpha}}.
\end{equation}
%
%
\subsubsection{$\alpha =d/2$ and $u\rightarrow u_c^{-}$}\label{sec3.4}
In this case the integral\ (\ref{eq2.6}) is logarithmically ultraviolet divergent. To fix the problem we start like in the preceding section and split the $k$-integral in the first line of\ (\ref{eq2.4}) into the sum of one integral over $\vert k\vert \ge\delta$ and an other integral over $\vert k\vert <\delta$, where $\delta >0$ is arbitrarily small. Let $\Delta_\delta (0)<+\infty$ denote the regularization of $\Delta (0)=+\infty$ with an infrared cutoff at $\vert k\vert =\delta$. Take $s>0$ real and make the change of variable $k=s^{1/\alpha}q$ in the low $\vert k\vert$ integral. One obtains
\begin{eqnarray}\label{eq2.9}
I(s)&\simeq &
I(0)+u_c s-\frac{\Delta_\delta (0)^2}{2}s^2 \\
&-&\frac{s^2}{\varepsilon}\int_{\vert q\vert <\delta /s^{2/d}}\left\lbrack
\frac{\varepsilon\tilde{C}(0)}{a\vert q\vert^{d/2}}-
\ln\left(1+\frac{\varepsilon\tilde{C}(0)}{a\vert q\vert^{d/2}}\right)\right\rbrack\, \frac{d^d q}{(2\pi)^d}, \nonumber
\end{eqnarray}
where the $q$-integral is now dominated by its ultraviolet behavior. Writing\ (\ref{eq2.9}) at second order in $s$, one gets
\begin{equation}\label{eq2.10}
I(s)\simeq I(0)+u_c s+Ks^{2}\ln(s)-\left\lbrack\Delta_\delta(0)^2+dK\ln(\delta)+O(1)\right\rbrack\frac{s^2}{2} ,
\end{equation}
with
\begin{equation}\label{eq2.11}
K=\frac{\varepsilon S_d\tilde{C}(0)^2}{(2\pi)^d d a^2},
\end{equation}
where $S_d$ is the unit sphere surface area in a $d$-dimensional space. Now, to deal with the logarithmic divergence of $\Delta_\delta(0)^2$ as $\delta$ tends to zero, we introduce a fixed $\eta >\delta$ small enough so that we can replace $g(k)$ with its expansion near $k=0$ for $\vert k\vert\le\eta$, and we write
\begin{eqnarray*}
\Delta_\delta(0)^2&=&\frac{1}{\varepsilon} \int_{|k|>\delta} {\frac{1}{g(k)^2}}\frac{d^{d}k}{(2 \pi)^d} \\
&=&\frac{\varepsilon \tilde{C}(0)^2 S_d}{(2\pi)^d a^2}\int_{\delta}^{\eta} \frac{dk}{k} +
\frac{1}{\varepsilon} \int_{|k|\ge\eta} {\frac{1}{g(k)^2}}\frac{d^{d}k}{(2 \pi)^d} \\
&=&\mathcal{T}-\frac{\varepsilon \tilde{C}(0)^2 S_d}{(2\pi)^d a^2} \ln(\delta),
\end{eqnarray*}
where
\begin{equation*}
\mathcal{T}=\frac{\varepsilon \tilde{C}(0)^2 S_d}{(2\pi)^d a^2} \ln(\eta)+\frac{1}{\varepsilon} \int_{|k|\ge\eta} {\frac{1}{g(k)^2}}\frac{d^{d}k}{(2 \pi)^d},
\end{equation*}
does not depend on $\delta$. It can easily be checked that $d\mathcal{T}/d\eta = 0$ too (as it should be, because $\Delta_\delta(0)^2$ cannot depend on an arbitrary quantity like $\eta$). For large $\vert k\vert$, $g(k)^{-2}$ behaves like $\tilde{C}(k)^2$ which is ultraviolet integrable, and $\mathcal{T}$ is finite. Using the expression\ (\ref{eq2.11}) for K, we see that the logarithmic divergence of $\Delta_\delta(0)^2$ compensate $K d \ln(\delta)$ exactly, yielding
\begin{equation}\label{bigt}
\lim_{\delta\rightarrow 0}\left\lbrack\Delta_\delta(0)^2+dK\ln(\delta)\right\rbrack =\mathcal{T} <+\infty ,
\end{equation}
and one gets
\begin{equation}\label{eq2.12}
I(s)=I(0)+u_c s+Ks^{2}\ln(s)+o(\vert s^{2}\ln(s)\vert).
\end{equation}
Since $I(s)$ is analytic in ${\mathbb C}^+$,\ (\ref{eq2.12}) holds not only for small $s$ in ${\mathbb R}^+$ but also for small $s$ in ${\mathbb C}^+$.

For $I(s)$ given by\ (\ref{eq2.12}) and $u<u_c$ close to $u_c$, the saddle point is solution to
\begin{equation*}
u=I^\prime (s(u))\simeq u_c+2Ks(u)\ln s(u),
\end{equation*}
which gives
\begin{equation*}
s(u)\simeq\frac{(u_c-u)/2K}{\vert\ln(u_c-u)/2K\vert},
\end{equation*}
and, to the same accuracy,
\begin{eqnarray}\label{eq2.13}
I(s(u))-us(u)&\simeq&\left\lbrack I(0)+u_c s(u)+Ks(u)^2\ln s(u)\right\rbrack
-\left\lbrack u_cs(u)+(u-u_c)s(u)\right\rbrack \nonumber \\
&=&I(0)+(u_c-u)s(u)+Ks(u)^2\ln s(u) \nonumber \\
&\simeq& I(0)+\frac{1}{4K}\, \frac{(u_c-u)^2}{\vert\ln(u_c-u)/2K\vert}.
\end{eqnarray}
Injecting\ (\ref{eq2.13}) into the right-hand side of\ (\ref{eq2.1}), one obtains
\begin{equation}\label{eq2.14}
\Xi(u)=I(0)+\frac{u}{\varepsilon\tilde{C}(0)}+\frac{1}{4K}\, \frac{(u_c-u)^2}{\vert\ln(u_c-u)/2K\vert}.
\end{equation}
%
%
\subsection{$\bm{\alpha <d}$ and $\bm{u>u_c}$}\label{sec3.5}
In this regime, the contribution of the $k=0$ mode becomes significant and the proper expression for $S(s,u,V)$ is the second line of\ (\ref{eq1.8}), which corresponds to $\sigma =1$ in\ (\ref{eq1.15}). The size of the $s$-domain contributing significantly to the Bromwich integral on the right-hand side of\ (\ref{eq1.15}) is then found to scale like $\sim 1/V(u-u_c)$ as $V\rightarrow +\infty$. For these $s$ one has $I(s)= I(0)+u_c s+o(1/V)$ and
\begin{equation}\label{eq2.15}
\int_{\mathcal L}s^{-1/\varepsilon}{\rm e}^{V\lbrack su-I(s)\rbrack}\frac{ds}{2i\pi}\sim
{\rm e}^{-VI(0)}\int_{\mathcal L}s^{-1/\varepsilon}\, {\rm e}^{V(u-u_c)s}\frac{ds}{2i\pi} =
\frac{{\rm e}^{-VI(0)}}{\lbrack \pi V(u-u_c)\rbrack^{1-1/\varepsilon}},
\end{equation}
from which it follows that\ (\ref{eq1.15}) reduces to
\begin{equation}\label{eq2.16}
\Xi(u)=I(0)+\frac{u}{\varepsilon\tilde{C}(0)}.
\end{equation}
%
%
\subsection{Conditions for a condensation of $\bm{\varphi}$}\label{sec3.6}
If $\alpha <d$, it can be seen immediately from\ (\ref{eq2.3}),\ (\ref{eq2.8}),\ (\ref{eq2.14}), and\ (\ref{eq2.16}) that $\Xi(u)$ is {\it not} analytic at $u=u_c$. This non analyticity indicates a phase transition of $\varphi$ at $u=u_c$. We will see in the next section that this phase transition corresponds actually to the onset of a condensation of $\varphi$, as mentioned at the end of Sec.\ \ref{sec2}.

On the other hand, if $\alpha\ge d$ one has $u_c =+\infty$ and the saddle point equation $u=I^\prime (s(u))$ has always a solution. No non analytic behavior of $\Xi(u)$, is to be expected in this regime.

\begin{figure}[htbp]
\begin{center}
\includegraphics [width=8cm] {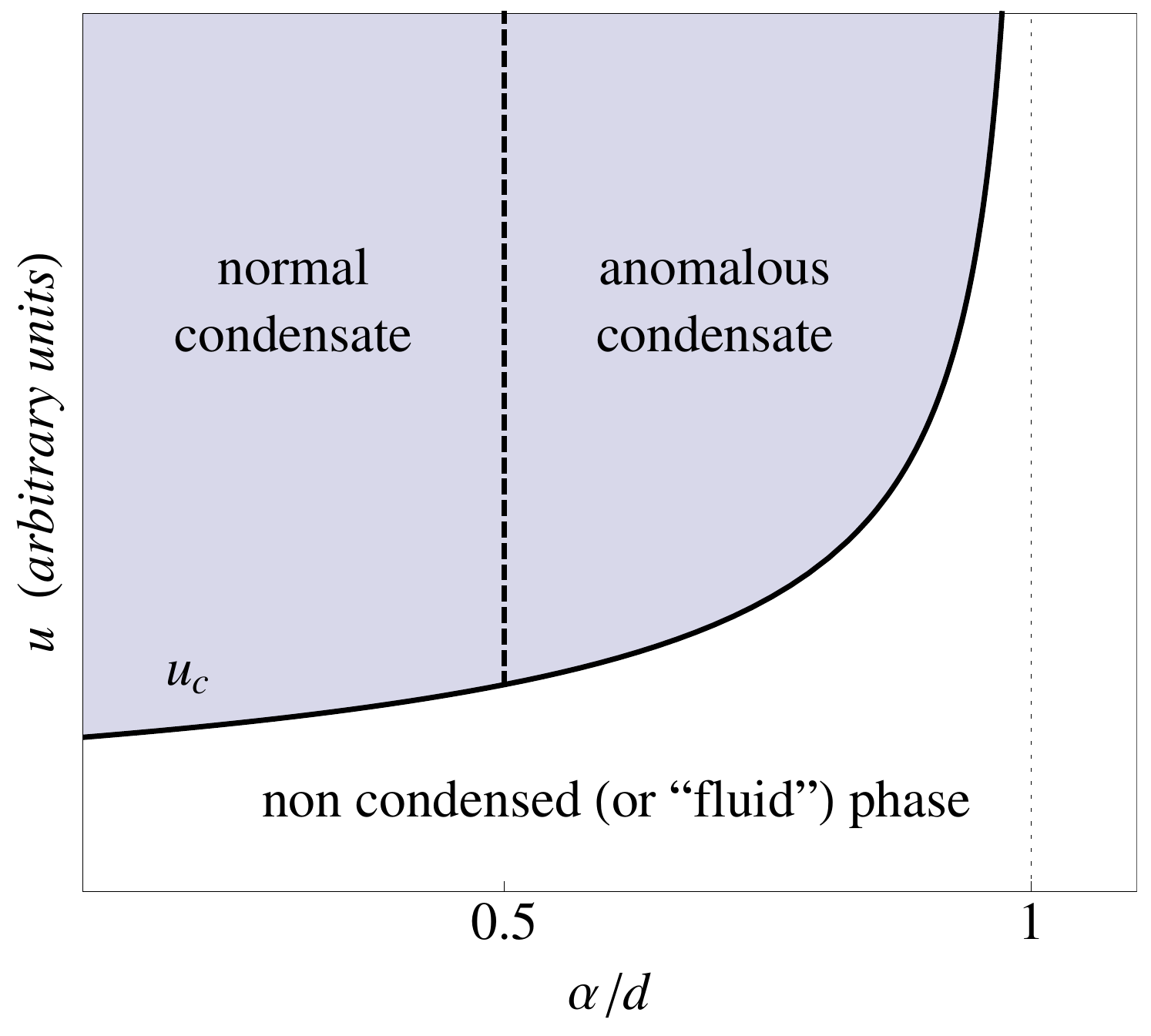}
\caption{\textsl{\it Schematic phase diagram in the $\alpha-u$ plane for a given dimension $d$, where $\alpha>0$ characterizes the small $k$ behavior of the Fourier transform of the covariance function: ${\tilde C}(k)\sim 1- a \vert k\vert^{\alpha}$ as $k\rightarrow 0$ (with $a>0$). There exists a critical line $u_c(\alpha,d)$ (solid line) separating the non condensed, or ``fluid", phase ($u<u_c(\alpha,d)$, white region) and the condensed phase ($u>u_c(\alpha,d)$, shaded region). The critical intensity $u_c(\alpha, d)$ diverges as $\alpha\to d$ from below (vertical dotted line). For $u>u_c(\alpha,d)$, there are two different types of condensates (separated by a vertical dashed line): a Gaussian distributed normal condensate for $0<\alpha<d/2$ and a non-Gaussian anomalous condensate for $d/2<\alpha<d$.} }
\label{fig1}
\end{center}
\end{figure}

The conditions for a Bose-Einstein condensation of $\varphi$ are thus : $\alpha <d$ and $u>u_c$. A schematic phase diagram in the $\alpha$-$u$ plane (for a given fixed dimension $d$) is presented in Fig. \ref{fig1}.
\bigskip
%
%
%
\section{The conditional spectral density of the field (given $\bm{u}$)}\label{sec4}
To specify the nature of the phase transition undergone by $\varphi$ at $u=u_c$ if $\alpha <d$, we can analyze the conditional spectral density of $\varphi$ for a given value of $u$ and see what happens when $u$ increases past $u_c$.

Since $u=\sum_{n\in{\mathbb Z}}m_n/V$ [see Eq.\ (\ref{cor1.4})], it is natural to define $u_n=m_n/V$ as the contribution of the mode $k=2\pi n/L$ to the intensity $u$. From the product measure\ (\ref{cor1.5}) one can easily write down the conditional joint pdf of the random variables $\lbrace u_n\rbrace$, conditioned to have a fixed sum $u=\sum_{n\in{\mathbb Z}}u_n$. One finds
\begin{equation}\label{cor3.1}
p\lbrack\lbrace u_n\rbrace\vert u\rbrack =
\frac{1}{p(u)}\, \delta\left(u-\sum_{i\in{\mathbb Z}^d}\right)\, \prod_{n\in{\mathbb Z}^d}p_n(u_n),
\end{equation}
where $p(u)$ is given by\ (\ref{eq1.7}) and, according to\ (\ref{cor1.5b}) and the change of variable $u_n=m_n/V$,
\begin{equation}\label{eq3.3}
p_n(u_n)=\frac{{\bm 1}_{\lbrace u_n>0\rbrace}}{\pi}\, 
\left(\frac{\pi V}{\varepsilon C_L(n)}\right)^{1/\varepsilon}\, \frac{1}{u_n^{1-1/\varepsilon}}\, \exp\left\lbrack -\frac{Vu_n}{\varepsilon C_L(n)}\right\rbrack ,
\end{equation}
with $\varepsilon =1,\, 2$ corresponding respectively to the complex and real cases. The conditional marginal density of a particular $u_n$ is then obtained from\ (\ref{cor3.1}) by integrating out all the other variables $u_{m\ne n}$. One gets
\begin{equation}\label{cor3.2}
p(u_n\vert u)=\frac{p_n(u_n) p_\bot(u-u_n)}{p(u)},
\end{equation}
where $p_\bot(u-u_n)$ is the probability density of the incomplete sum $\sum_{m\ne n}u_m =u-u_n$ (with given $u$ and $u_n$). Since the $u_m$ are independent random variables, $p_\bot(u-u_n)$ has the same functional form as $p(u)$ in Eq.\ (\ref{eq1.7}), except that $S(s,u,V)$ is now replaced with
\begin{equation}\label{eq3.5}
S_n(s,u,V)=su-\frac{1}{\varepsilon V}
\sum_{m\ne n}\ln\left\lbrack 1-\frac{C_L(m)}{C_L(0)}+\varepsilon sC_L(m)\right\rbrack ,
\end{equation}
namely
\begin{equation}\label{eq3.4}
p_\bot(u-u_n)=V{\rm e}^{-\frac{V(u-u_n)}{\varepsilon C_L(n)}}\, \int_{\mathcal{L}}
{\rm e}^{VS_n(s,u-u_n,V)}\frac{ds}{2i\pi}.
\end{equation}
Using the conditional marginal distribution $p(u_n\vert u)$ given by\ (\ref{cor3.2}), we can define the conditional expectation
\begin{equation}\label{eq3.2}
\mathbb{E}(u_n\vert u)=\int_0^{+\infty}u_n p(u_n\vert u)\, du_n
=\frac{1}{p(u)}\int_0^{+\infty}u_np_n(u_n)p_\bot(u-u_n)\, du_n.
\end{equation}
It is easy to check that $\sum_n\mathbb{E}(u_n\vert u) =u$, as it should be. Physically, $\mathbb{E}(u_n\vert u)$ represents the average fraction of the total intensity $u$ carried by the $n$-th mode.

The conditional spectral density, $\tilde{C}_u(k)=\tilde{C}_u(\vert k\vert)$, is subsequently defined in the thermodynamic limit by requiring $\tilde{C}_u(k)\, d^dk/(2\pi)^d$ to represent the average fraction of the total intensity $u$ carried by all the modes that lie within an infinitesimal spectral volume $d^dk$ around $k$. According to this definition, one has
\begin{equation}\label{eq3.1}
\tilde{C}_u(k)=-\frac{(2\pi)^d}{S_d\vert k\vert^{d-1}}\, \frac{d}{d\vert k\vert}
\lim_{V\rightarrow +\infty}\sum_{\vert n\vert\ge\vert k\vert L/2\pi}\mathbb{E}(u_n\vert u),
\end{equation}
which is obviously normalized to $\int\tilde{C}_u(k)\, d^dk/(2\pi)^d=u$, as a straightforward consequence of $\sum_n\mathbb{E}(u_n\vert u) =u$.

Before proceeding with the determination of $\tilde{C}_u(k)$ in the different regimes found in Sec.\ \ref{sec3}, we first put $\mathbb{E}(u_n\vert u)$ into a form more suitable to an asympotic estimate in the thermodynamic limit. From the expressions\ (\ref{eq1.7}),\ (\ref{eq3.3}),\ (\ref{eq3.4}), and the Laplace transform
\begin{equation*}
\int_0^{+\infty}u_n^{1/\varepsilon}{\rm e}^{-\frac{V}{\varepsilon}\left(\frac{1}{C_L(n)}-\frac{1}{C_L(0)}+\varepsilon s\right)\, u_n}du_n
=\frac{\pi^{1-1/\varepsilon}}{\varepsilon\lbrack (V/\varepsilon)(C_L(n)^{-1}-C_L(0)^{-1}+\varepsilon s)\rbrack^{1+1/\varepsilon}},
\end{equation*}
(where we have written $\Gamma(1+1/\varepsilon)=\pi^{1-1/\varepsilon}/\varepsilon$, valid for $\varepsilon =1,\, 2$), one finds after some straightforward algebra,
\begin{equation}\label{eq3.6}
\mathbb{E}(u_n\vert u)=\left(\int_{\mathcal{L}}\frac{C_L(n)\, {\rm e}^{VS(s,u,V)}}{1-C_L(n)/C_L(0)+\varepsilon sC_L(n)}\, \frac{ds}{2i\pi}\right)
\left(V\int_{\mathcal{L}}{\rm e}^{VS(s,u,V)}\frac{ds}{2i\pi}\right)^{-1}.
\end{equation}
%
%
\subsection{$\bm{\alpha <d}$ and $\bm{u< u_c}$, or $\bm{\alpha\ge d}$: no condensation}\label{sec4.1}
In this regime, the proper expression for $S(s,u,V)$ in the large $V$ limit is the first line of\ (\ref{eq1.8}) and there always is one saddle point $s(u)>0$, solution to $u=I^\prime (s(u))$. This saddle point determines the Bromwich integrals on the right-hand side of\ (\ref{eq3.6}) which reduces to
\begin{equation}\label{eq3.7}
\lim_{V\rightarrow +\infty}\sum_{\vert n\vert\ge\vert k\vert L/2\pi}\mathbb{E}(u_n\vert u)=
\int_{\vert q\vert\ge\vert k\vert}\frac{\tilde{C}(q)}{1-\tilde{C}(q)/\tilde{C}(0)+\varepsilon s(u)\tilde{C}(q)}\, \frac{d^dq}{(2\pi)^d},
\end{equation}
and, according to the definition\ (\ref{eq3.1}),
\begin{equation}\label{eq3.8}
\tilde{C}_u(k)=\frac{\tilde{C}(k)}{1-\tilde{C}(k)/\tilde{C}(0)+\varepsilon s(u)\tilde{C}(k)}.
\end{equation}
Replacing $I(s)$ with its expression\ (\ref{eq1.9}) in the saddle point equation $u=I^\prime (s(u))$ and using\ (\ref{eq3.8}), one obtains $\int\tilde{C}_u(k)\, d^dk/(2\pi)^d=u$, which means that\ (\ref{eq3.8}) gives the complete distribution of $u$ among {\it all} the Fourier modes. It is clear from this result that, in the limit $V\rightarrow+\infty$, the contribution of each given mode to $u$ is infinitesimal and no condensate is present in this regime.

Note that for $u$ close to $u_c$, (or $u$ large enough, if $u_c =+\infty$), $s(u)$ is small and\ (\ref{eq3.8}) can be approximated by
\begin{equation}\label{eq3.9}
\begin{array}{lr}
\tilde{C}_u(k)\simeq\frac{\tilde{C}(k)}{1-\tilde{C}(k)/\tilde{C}(0)}&(\vert k\vert^\alpha\gg\varepsilon s(u)/a), \\
\tilde{C}_u(k)\simeq\frac{1}{a\vert k\vert^\alpha +\varepsilon s(u)}&(\vert k\vert^\alpha\lesssim\varepsilon s(u)/a),
\end{array}
\end{equation}
which depends on $u$ in the vicinity of $k=0$ only.
%
%
\subsection{$\bm{\alpha <d}$ and $\bm{u>u_c}$: condensation}\label{sec4.2}
As mentioned in Sec.\ \ref{sec3.5}, the proper expression for $S(s,u,V)$ in this regime is the second line of\ (\ref{eq1.8}). It follows that the size of the contributing $s$ on the right-hand side of\ (\ref{eq3.6}) goes to zero like $\sim 1/V(u-u_c)$ as $V\rightarrow +\infty$, and for $k\ne 0$ one has
\begin{equation}\label{eq3.10}
\lim_{V\rightarrow +\infty}\sum_{\vert n\vert\ge\vert k\vert L/2\pi}\mathbb{E}(u_n\vert u)=
\int_{\vert q\vert\ge\vert k\vert}\frac{\tilde{C}(q)}{1-\tilde{C}(q)/\tilde{C}(0)}\, \frac{d^dq}{(2\pi)^d},
\end{equation}
which yields
\begin{equation}\label{eq3.11}
\tilde{C}_u(k)=\frac{\tilde{C}(k)}{1-\tilde{C}(k)/\tilde{C}(0)}.
\end{equation}
Replacing $g(k)$ with its expression\ (\ref{eq1.10}) in the definition\ (\ref{eq1.11}) of $u_c$ and using\ (\ref{eq3.11}), one obtains $\int_{k\ne 0}\tilde{C}_u(k)\, d^dk/(2\pi)^d=u_c<u$. The normalization $\int\tilde{C}_u(k)\, d^dk/(2\pi)^d=u$ demands then that the missing part $u-u_c$ accumulate into the $k=0$ mode. Let us check it explicitely. For $k=0$ the numerator on the right-hand side of\ (\ref{eq3.6}) is proportional to
\begin{equation}\label{eq3.12}
\frac{1}{\varepsilon}\int_{\mathcal L}s^{-(1+1/\varepsilon)}{\rm e}^{V\lbrack su-I(s)\rbrack}\frac{ds}{2i\pi}\sim
\frac{{\rm e}^{-VI(0)}}{\varepsilon}\int_{\mathcal L}s^{-(1+1/\varepsilon)}\, {\rm e}^{V(u-u_c)s}\frac{ds}{2i\pi} =
\frac{{\rm e}^{-VI(0)}\lbrack V(u-u_c)\rbrack^{1/\varepsilon}}{\pi^{1-1/\varepsilon}},
\end{equation}
with $\varepsilon =1,\, 2$. And since the Bromwich integral in the denominator is proportional to\ (\ref{eq2.15}) with the same coefficient of proportionality, one obtains
\begin{equation}\label{eq3.13}
\lim_{V\rightarrow +\infty}\mathbb{E}(u_0\vert u)=u-u_c,
\end{equation}
as expected. Combining Equations\ (\ref{eq3.10}) and\ (\ref{eq3.13}), one finds that the spectral density\ (\ref{eq3.11}) can be prolongated to $k=0$, in the distributional sense, as
\begin{equation}\label{eq3.14}
\tilde{C}_u(k)=(2\pi)^d (u-u_c)\, \delta(k)+\frac{\tilde{C}(k)}{1-\tilde{C}(k)/\tilde{C}(0)}{\bm 1}_{\lbrace k\ne 0\rbrace}.
\end{equation}
The condensation into the $k=0$ mode manifests itself in the presence of the $\delta$-function on the right-hand side of\ (\ref{eq3.14}). The sudden appearance of a condensate as $u$ increases past $u_c$ is a clear manifestation of the non analyticity of $\Xi(u)$ [or, equivalently, of $p(u)$] at $u=u_c$. 
%
%
\section{The structure of the condensate ($\bm{\alpha <d}$ and $\bm{u>u_c}$)}\label{sec5}
In this section, we investigate the structure of the condensate in more detail in terms of the asymptotic behavior as $V\rightarrow +\infty$ of the conditional marginal density of the zero mode $u_0$ (given $u$),
\begin{equation}\label{eq5.1}
p(u_0|u)=\frac{p_0(u_0) p_{\bot}(u-u_0)}{p(u)},
\end{equation}
for $u>u_c$. Using\ (\ref{eq1.7}),\ (\ref{eq3.3}), and\ (\ref{eq3.4}) with $n=0$ in\ (\ref{eq5.1}), one obtains
\begin{equation}\label{eq5.2}
p(u_0|u)=\frac{V}{(\pi V u_0)^{1-1/\varepsilon}}
\left(\int_{\mathcal{L}}s^{1/\varepsilon}{\rm e}^{V\lbrack S(s,u,V)-u_0s\rbrack}\frac{ds}{2i\pi}\right)
\left(\int_{\mathcal{L}}{\rm e}^{VS(s,u,V)}\frac{ds}{2i\pi}\right)^{-1}.
\end{equation}
Since $u>u_c$, the proper asymptotic expression for $S(s,u,V)$ in the denominator of\ (\ref{eq5.2}) is the second line of\ (\ref{eq1.8}), and from\ (\ref{eq2.15}) one gets
\begin{eqnarray}\label{eq5.3}
p(u_0\vert u)&\sim& V\lbrack\varepsilon\tilde{C}(0)f(V)\rbrack^{1/\varepsilon}
\left(\frac{u-u_c}{u_0}\right)^{1-1/\varepsilon}{\rm e}^{VI(0)} \nonumber \\
&\times&\int_{\mathcal{L}}s^{1/\varepsilon}{\rm e}^{V\lbrack S(s,u,V)-u_0s\rbrack}\frac{ds}{2i\pi}.\ \ \ \ (V\rightarrow +\infty)
\end{eqnarray}
The asymptotic behavior of the Bromwich integral on the right-hand side of\ (\ref{eq5.3}) depends on the considered regime non trivially, as we will now see.
\subsection{$\bm{\alpha < d/2}$}\label{sec5.1}
Since $u_0$ is expected to be close to $u-u_c$, the integral in\ (\ref{eq5.3}) is expected to be determined by the contribution of the small $s$ region. The actual size of this region determines which asymptotic expression\ (\ref{eq1.8}) is to be chosen. Leaving this choice open for the moment and using\ (\ref{eq5.3}), one obtains
\begin{eqnarray}\label{eq5.4}
p(u_0\vert u)&\sim& V\lbrack\varepsilon\tilde{C}(0)f(V)\rbrack^{(1-\sigma)/\varepsilon}
\left(\frac{u-u_c}{u_0}\right)^{1-1/\varepsilon}{\rm e}^{VI(0)} \nonumber \\
&\times&\int_{\mathcal L}s^{(1-\sigma)/\varepsilon}{\rm e}^{V\lbrack s(u-u_0)-I(s)\rbrack}\frac{ds}{2i\pi},\ \ \ \ (V\rightarrow +\infty)
\end{eqnarray}
where $\sigma =0$ (resp. $\sigma =1$) if the first (resp. second) line of\ (\ref{eq1.8}) is the right choice. Since $\Delta (0) <+\infty$ if $\alpha < d/2$, one can write, for small $s$,
\begin{equation}\label{eq5.5}
I(s)=I(0)+u_c s-\frac{1}{2}\Delta(0)^2 s^2+o(\vert s\vert^2).
\end{equation}
Injecting\ (\ref{eq5.5}) into the right-hand side of\ (\ref{eq5.4}) and writing
\begin{equation}\label{eq5.6}
s=\frac{u_0-u+u_c}{\Delta(0)^2}+t,
\end{equation}
one gets
\begin{eqnarray}\label{eq5.7}
p(u_0\vert u)&\sim& V\lbrack\varepsilon\tilde{C}(0)f(V)\rbrack^{(1-\sigma)/\varepsilon}
\left(\frac{u-u_c}{u_0}\right)^{1-1/\varepsilon}{\rm e}^{-\frac{V(u_0-u+u_c)^2}{2\Delta(0)^2}} \nonumber \\
&\times&\int_{\mathcal L}\left(\frac{u_0-u+u_c}{\Delta(0)^2}+t\right)^{(1-\sigma)/\varepsilon}
{\rm e}^{\frac{V}{2}\Delta(0)^2 t^2}\frac{dt}{2i\pi}.\ \ \ \ (V\rightarrow +\infty)
\end{eqnarray}
The value of $\sigma$ can now be determined by comparing, for the contributing $s$, the speed at which $sC_L(0)$ and $1-C_L(\vert n\vert =1)/C_L(0)$ in\ (\ref{eq1.6}) tend to zero as $V\rightarrow +\infty$. For instance, a rapidly decreasing $sC_L(0)$ going to zero faster than $1-C_L(\vert n\vert =1)/C_L(0)$ is equivalent to letting $s\rightarrow 0$ first, and then $V\rightarrow +\infty$. In this case, the asymptotic behavior of $S(s,u,V)$ is given by the second line of\ (\ref{eq1.8}) and one must take $\sigma =1$. In the opposite case where $1-C_L(\vert n\vert =1)/C_L(0)$ goes to zero faster than $sC_L(0)$, one must take $\sigma =0$. So, we need to estimate the typical size of the contributing $s$. Because of the exponential in the first line of\ (\ref{eq5.7}), the distribution of $u_0$ is peaked around $u_0=u-u_c$ with a width $\sim 1/\sqrt{V}$. Because of the exponential in the second line of\ (\ref{eq5.7}), the contibuting $t$ are in a small region around $t=0$ with a width $\sim 1/\sqrt{V}$. Thus, according to\ (\ref{eq5.6}), the small $s$ contributing to\ (\ref{eq5.4}) for a typical $u_0$ behave like $1/\sqrt{V}=1/L^{d/2}$ as $V\rightarrow +\infty$, and since $\alpha <d/2$, one finds
\begin{equation*}
\lim_{V\rightarrow +\infty}\left\vert\frac{sC_L(0)}{1-C_L(\vert n\vert =1)/C_L(0)}\right\vert
=\frac{\tilde{C}(0)}{a(2\pi)^\alpha}\lim_{L\rightarrow +\infty}sL^\alpha =0,
\end{equation*}
yielding $\sigma =1$, as explained above. The integral in\ (\ref{eq5.7}) reduces then to a simple Gaussian integral. Using $(u-u_c)/u_0\rightarrow 1$ as $V\rightarrow +\infty$ (again, because the distribution is peaked around $u_0=u-u_c$ with a width $\sim 1/\sqrt{V}$), one obtains
\begin{equation}\label{eq5.8}
p(u_0\vert u)\sim\sqrt{\frac{V}{2\pi}}\frac{{\rm e}^{-\frac{V(u_0-u+u_c)^2}{2\Delta(0)^2}}}{\Delta(0)}.\ \ \ \ (V\rightarrow +\infty)
\end{equation}
In this regime, the part of $u$ which is in the condensate is Gaussian distributed around $u-u_c$ with a width $\sim 1/\sqrt{V}$. We will call such a condensate a ``normal condensate".
\subsection{$\bm{d/2\le\alpha<d}$ : statement of the problem and solution for $\bm{d=1}$}\label{sec5.2}
The most natural way of proceeding to the case $d/2<\alpha<d$ would be to start again from Expression\ (\ref{eq5.4}) and determine the asymptotic behavior of the Bromwich integral by replacing $I(s)$ with the small $s$ expansion\ (\ref{eq2.5}), instead of\ (\ref{eq5.5}). Unfortunately, this line turns out to be a dead end. The size of the $s$ contributing to\ (\ref{eq5.4}) for a typical $u_0$ is found to behave like $1/V^{\alpha /d}=1/L^{\alpha}$ for large $V$. It follows that $L^\alpha s\rightarrow O(1)$ as $V\rightarrow +\infty$ for the contributing $s$, and no definite value can be attributed to $\sigma$.

As we will see in the following, this unexpected problem is to be attributed to the fact that the asymptotics\ (\ref{eq1.8}) are not correct if $L^\alpha s\rightarrow O(1)$ or $L^\alpha s\rightarrow 0$ too slowly (e.g. logarithmically) as $V\rightarrow +\infty$. A more careful, and finer, analysis of the asymptotic behavior of\ (\ref{eq1.6}) is therefore needed to determine the one of $p(u_0\vert u)$ from the expression\ (\ref{eq5.3}).

We have been able to carry out such an analysis successfully for $d=1$, taking advantage of the Euler-Maclaurin summation formula to control\ (\ref{eq1.6}) and  obtain $p(u_0\vert u)$ as a Laplace integral which can be computed numerically. The result is the existence of an ``anomalous condensate"  if $d/2\le\alpha < d$ and $u>u_c$.
\subsubsection{Rewriting $S(s,u,V)$ using the Euler-Maclaurin summation formula}
From the Euler-Maclaurin summation formula at lowest order,
\begin{equation*}
\sum^{+\infty}_{n=1} f(n)=-\frac{f(0)}{2}+\int_{0}^{+\infty} f(t) dt + \int_{0}^{+\infty} P_{1}(t)f^{'}(t) dt,
\end{equation*}
where $P_{1}(t)=t-n-1/2$ for $n<t<n+1$, one gets
\begin{eqnarray} \label{eq5.9} 
\sum^{+\infty}_{-\infty} \ln\left[1-\frac{C_L(n)}{C_L(0)}+\varepsilon s C_L(n)  \right]&=&\int_{-\infty}^{+\infty} \ln\left[1-\frac{C_L(t)}{C_L(0)} +\varepsilon s C_L(t) \right] dt \nonumber \\
&+&2\left[\varepsilon s-\frac{1}{C_L(0)} \right] \int_{0}^{+\infty} \frac{P_1(t) C^{'}_{L}(t)}{1-\frac{C_{L}(t)}{C_{L}(0)}+ \varepsilon s C_{L}(t)} dt.
\end{eqnarray}
Making the change of variable $t=kL/2\pi$, one obtains
\begin{eqnarray} \label{eq5.10}
\sum^{+\infty}_{-\infty} \ln\left[1-\frac{C_L(n)}{C_L(0)}+\varepsilon s C_L(n) \right]&=&L\int_{-\infty}^{+\infty} \ln\left[1-\frac{\tilde{C}_L(k)}{\tilde{C}_L(0)} +\varepsilon s \tilde{C}_L(k) \right] \frac{dk}{2\pi} \nonumber \\
&+&2\left[\varepsilon s-\frac{1}{\tilde{C}_{L}(0)} \right] \int_{0}^{+\infty} \frac{P_1(\frac{kL}{2 \pi}) \tilde{C}^{'}_{L}(k)}{1-\frac{\tilde{C}_{L}(k)}{\tilde{C}_{L}(0)}+ \varepsilon s \tilde{C}_{L}(k)} dk,
\end{eqnarray}
where we have written $C_L(kL/2\pi)=\tilde{C}_{L}(k)$ and $C_L^\prime (kL/2\pi)=(2\pi /L)\, \tilde{C}_{L}^{'}(k)$. Let $R(s,L)$ denote the second term on the right hand side of (\ref{eq5.10}). When $s\rightarrow 0$, $R(s,L)$ is logarithmically infrared divergent. To deal with this divergence, we rewrite $R(s,L)$ as
\begin{equation} \label{eq5.11}
R(s,L)=\ln{\left[\varepsilon s \tilde{C}_L(0)\right]} +2\left[\varepsilon s-\frac{1}{\tilde{C}_{L}(0)} \right] \int_{0}^{+\infty} \frac{[P_1(\frac{kL}{2 \pi})+1/2] \tilde{C}^{'}_{L}(k)}{1-\frac{\tilde{C}_{L}(k)}{\tilde{C}_{L}(0)}+ \varepsilon s \tilde{C}_{L}(k)} dk.
\end{equation}
For $d=1$, it follows from\ (\ref{eq1.6}),\ (\ref{eq5.10}), and\ (\ref{eq1.9}) that
\begin{equation}\label{eq5.12}
S(s,u,L)\sim su-I(s)-\frac{R(s,L)}{\varepsilon L}\ \ \ \ (L\rightarrow +\infty).
\end{equation}
Our objective is now to determine the behavior of\ (\ref{eq5.12}) as $s\rightarrow 0$ with $L^\alpha s\rightarrow O(1)$, and use the result in the Bromwich integral on the right-hand side of\ (\ref{eq5.3}) to get the asymptotic behavior of $p(u_0\vert u)$ as $L\rightarrow +\infty$ for $d=1$, $1/2<\alpha<1$, and $u>u_c$.
\subsubsection{Determination of $p(u_0\vert u)$ for $d=1$, $1/2<\alpha<1$, and $u>u_c$} 
After some careful calculations detailed in Appendix\ \ref{app2}, one finds that in the limits $L\rightarrow +\infty$, $s\rightarrow 0$, and $L^\alpha s\le O(1)$, $R(s,L)$ behaves as
\begin{eqnarray} \label{eq5.13}
R(s,L)\sim R(sL^\alpha)&=&\ln\left(\frac{\varepsilon \tilde{C}_{L}(0) sL^\alpha}{a}\right)
+2\phi\left(\frac{\varepsilon\tilde{C}_{L}(0)}{a} (L/2\pi)^\alpha s\right) \nonumber \\
&-&\frac{\alpha}{\pi} \Gamma(1+1/\alpha) \Gamma(1-1/\alpha)\left(\frac{\varepsilon \tilde{C}_{L}(0) sL^\alpha}{a}\right)^{1/\alpha} ,
\end{eqnarray}
with
\begin{equation}\label{eq5.14}
\phi(t)=\zeta(\alpha)\, t +\sum_{n=1}^{+\infty}\left[\ln\left(1+\frac{t}{n^\alpha}\right) -\frac{t}{n^\alpha} \right ] ,
\end{equation}
where $\zeta(\cdot)$ is the Riemann zeta function. The second line of\ (\ref{eq1.8}) corresponds to the first term on the right-hand side of\ (\ref{eq5.13}) only, with $f(V)=L^\alpha /a$. Injecting\ (\ref{eq5.13}) into\ (\ref{eq5.12}), replacing $I(s)$ with the small $s$ expansion\ (\ref{eq2.5}), and introducing the rescaled variables
\begin{eqnarray*}
&&t=\frac{\varepsilon \tilde{C}(0)}{ia} \left(\frac{L}{2\pi}\right)^\alpha s, \\
&&X=\frac{(2\pi)^\alpha a}{\varepsilon \tilde{C}(0)}\, L^{1-\alpha}(u_0-u+u_c),
\end{eqnarray*}
one obtains $p(u_0|u)$ for $u_0$ close to $u-u_c$ as
\begin{equation} \label{eq5.15}
p(u_0|u)\sim\frac{(2\pi)^\alpha a}{\varepsilon \tilde{C}(0)}\, L^{1-\alpha} \int_{-\infty}^{+\infty}{\rm e}^{-iXt-\frac{2}{\varepsilon} \phi(i t)} \frac{dt}{2\pi}.
\ \ \ \ (L\rightarrow +\infty)
\end{equation}
Note that by the identity
\begin{equation*}
\int_{0}^{\infty} \left\lbrack{\frac{1}{t^\alpha}-\ln{\left(1+\frac{1}{t^\alpha}\right)}}\right\rbrack dt=
-\alpha \Gamma\left(1+\frac{1}{\alpha}\right)\Gamma\left(1-\frac{1}{\alpha}\right),
\end{equation*}
the term in the second line of\ (\ref{eq5.13}) exactly cancels out the term $-Js^{1/\alpha}$ from\ (\ref{eq2.5}). The simple asymptotics given by the second line of\ (\ref{eq1.8}) misses this cancellation.
\bigskip
\begin{figure}[htbp]
\begin{center}
\includegraphics [width=8cm] {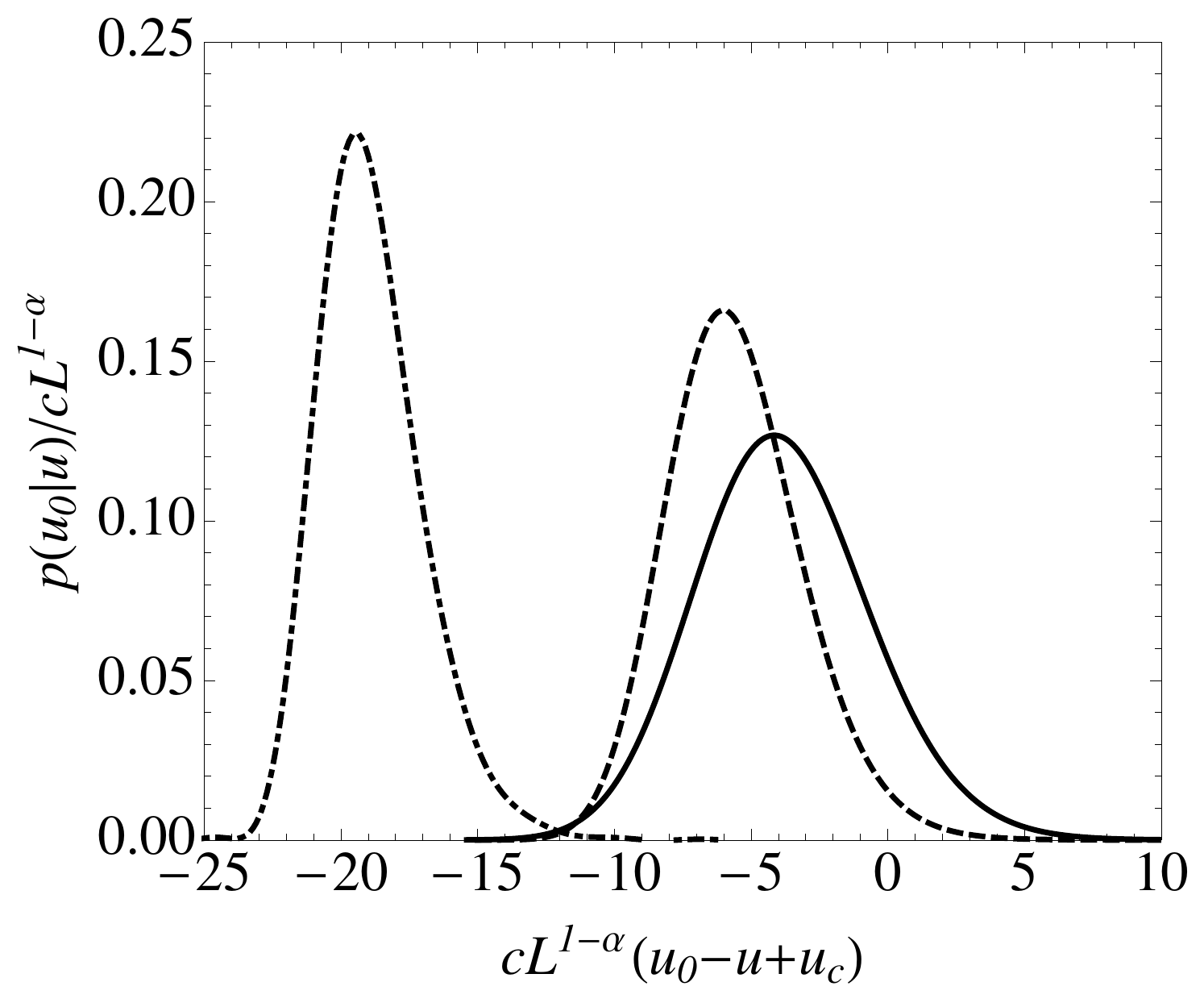}
\caption{\textsl{\it Rescaled $p(u_0|u)$ vs. the rescaled variable $X$ with $c=(2\pi)^\alpha a/\varepsilon \tilde{C}(0)$ (see the text for the definition of $X$), for $d=1$ and $\alpha=0.6$ (solid line), $\alpha=0.7$ (dashed line), and $\alpha=0.9$ (dot-dashed line).} }
\label{fig2}
\end{center}
\end{figure}

The $t$-integral on the right-hand side of\ (\ref{eq5.15}) can be computed numerically for fixed $X$. Then, varying $X$ one obtains the kind of curves shown in Fig.\ \ref{fig2}. The non-Gaussian nature of the condensate in this regime is easily seen for, e.g., $\alpha=0.7$ and $\alpha=0.9$. More quantitatively, the skewness increases from $0.22$ for $\alpha=0.6$, to $0.40$ for $\alpha=0.7$, to $0.67$ for $\alpha=0.9$. Similarly, the excess kurtosis increases from $0.11$ for $\alpha=0.6$, to $0.39$ for $\alpha=0.7$, to $0.81$ for $\alpha=0.9$. Note that, as a function of $u_0$, the probability distribution $p(u_0|u)$ always tends to a delta function located at $u_0=u-u_c+0^{-}$ as $L\rightarrow +\infty$. To see the structure of the condensate it is necessary to zoom in on the delta function, which is achieved by using the rescaled variable $X$ instead of $u_0$, and rescaling $p(u_0|u)$ accordingly. The shifting of the peak of $p(u_0|u)$ to the left with increasing $\alpha$, observed in Fig.\ \ref{fig2}, must be attributed to the fact that $\alpha =d$ is the limit for Bose-Einstein condensation (here $d=1$). The closer to $\alpha =d$ one is, the larger $L$ must be for the peak of $p(u_0|u)$ to eventually get close to $u_0 = u-u_c$ by less than a given finite amount. The necessary $L$ diverges in the limit $\alpha\rightarrow d^{-}$, which is the precursor of the destruction of the condensate at $\alpha = d$.

In summary, at least for $d=1$ where we are able to analyse the distribution $p(u_0|u)$ accurately, it turns out that the anomalous condensate in our case has a much more complex structure than the relatively simple form of the anomalous condensate found in the homogeneous mass transport model\ \cite{MEZ,EMZ}. 
\subsubsection{Determination of $p(u_0\vert u)$ for $d=1$, $\alpha =1/2$, and $u>u_c$}
The existence of an anomalous condensate for $1/2<\alpha <1$ raises the question of the structure of the condensate at the transition $\alpha =1/2$ between the normal and anomalous regimes. In this case, the behavior of $R(s,L)$ in the limits $L\rightarrow +\infty$, $s\rightarrow 0$, and $L^{1/2} s\le O(1)$, is found to be given by
\begin{eqnarray}\label{eq5.16}
R(s,L)\sim R(L^{1/2} s)&=&\ln\left(\frac{\varepsilon\tilde{C}(0) L^{1/2}s}{a}\right)
+2\psi\left(\frac{\varepsilon\tilde{C}(0)L^{1/2}s}{a\sqrt{2\pi}}\right)\nonumber \\
&+&\frac{\varepsilon LKs^2}{2}\left\lbrack 1-2\ln\left(\frac{\varepsilon \tilde{C}(0)L^{1/2}s}{a\sqrt{2\pi}}\right)\right\rbrack,
\end{eqnarray}
with $K=\varepsilon\tilde{C}(0)^2/(\pi a^2)$ [see Eq.\ (\ref{eq2.11})], and
\begin{equation}\label{eq5.17}
\psi(t)=\zeta\left(\frac{1}{2}\right) t -\frac{\gamma t^2}{2}
+\sum_{n=1}^{+\infty}\left\lbrack\ln\left(1+\frac{t}{\sqrt{n}}\right)-\frac{t}{\sqrt{n}}+\frac{t^2}{2n} \right\rbrack ,
\end{equation}
where $\gamma$ is the Euler-Mascheroni constant. The interested reader is referred to Appendix\ \ref{app2} for the details of the calculation. Injecting\ (\ref{eq5.16}) into\ (\ref{eq5.12}) with
\begin{equation}\label{eq5.18}
I(s)\simeq I(0)+u_c s+Ks^{2}\ln\left(\frac{\varepsilon\tilde{C}(0)s}{a}\right)
-(\mathcal{T}+K)\frac{s^2}{2} ,
\end{equation}
[see Eq.\ (\ref{eq2.10}) with $O(1)$ made explicit, and Eq.\ (\ref{bigt})], and introducing the rescaled variable
\begin{equation*}
t=\frac{\tilde{C}(0)}{ia}\, \left(\frac{\varepsilon L\ln L}{2\pi}\right)^{1/2}s,
\end{equation*}
one obtains $p(u_0|u)$ for $u_0$ close to $u-u_c$ as
\begin{eqnarray} \label{eq5.19}
p(u_0|u)&\sim&\frac{1}{\pi\sqrt{2}\, \Delta_L} \int_{-\infty}^{+\infty}{\rm e}^{-i\frac{\sqrt{2}}{\Delta_L}\, (u_0-u+u_c)t -t^2} dt \nonumber \\
&=&\frac{1}{\sqrt{2\pi}}\, \frac{{\rm e}^{-\frac{(u_0-u+u_c)^2}{2\Delta_L^2}}}{\Delta_L},\ \ \ \ (L\rightarrow +\infty)
\end{eqnarray}
where
\begin{equation}\label{eq5.20}
\Delta_L=\frac{\varepsilon\tilde{C}(0)}{a}\left(\frac{\ln L}{\pi L}\right)^{1/2}.
\end{equation}
Like in the normal regime ($\alpha <1/2$), the part of $u$ which is in the condensate is Gaussian distributed around $u-u_c$ but with anomalously large fluctuations scaling as $\sqrt{\ln(L)/L}$, instead of $1/\sqrt{L}$ for the normal condensate.

It is interesting to notice that the size of the $s$ contributing to\ (\ref{eq5.3}) for a typical $u_0$ in this regime is found to behave like $1/\sqrt{L\ln L}$ for large $L$. It follows that $L^{1/2} s\sim 1/\sqrt{\ln L}\rightarrow 0$ as $L\rightarrow +\infty$ for the contributing $s$, which suggests intuitively that the second line of\ (\ref{eq1.8}) might be the right asymptotics to use. This is not the case. Like in the previous regime, Expression\ (\ref{eq5.19}) results from non trivial cancellations between terms of\ (\ref{eq5.16}) and\ (\ref{eq5.18}) that are not accounted for by the simple asympotics\ (\ref{eq1.8}).
%
%
\section{Summary and perspectives}\label{sec6}
In this paper, we have investigated the concentration properties of a Gaussian random field $\varphi$ in the thermodynamic limit, $V\rightarrow +\infty$ with fixed $u=\|\varphi\|_2^2/V$, where $V$ is the volume of the box containing the field. Considering a wide class of fields for which the spectral density behaves like $\tilde{C}(k)\sim\tilde{C}(0)\left(1-a\vert k\vert^\alpha\right)$ for small $k$ (with $a,\, \alpha >0$), we have found different regimes depending on the relative values of $\alpha$ and the space dimension $d$.

If $\alpha <d$ and $u>u_c$, where $u_c$ is given by Eq.\ (\ref{eq1.11}), the concentration of $\varphi$ discussed in\ \cite{MD} and\ \cite{MC} turns into a Bose-Einstein condensation onto the $k=0$ mode as the thermodynamic limit is taken. If $\alpha <d/2$, the condensate is normal in the sense that the part of $u$ which is in the condensate is Gaussian distributed around $u-u_c$ with normal fluctuations scaling as $1/\sqrt{V}$. If $d/2<\alpha <d$, the condensate is not Gaussian distributed (``anomalous" condensate). A detailed semi-analytic analysis of the structure of the anomalous condensate has been performed for $d=1$. Extending this one-dimensional analysis to the transition between normal and anomalous condensations, we have found that the condensate at the transition is still Gaussian distributed but with anomalously large fluctuations scaling as $\sqrt{\ln(L)/L}$.

If $\alpha\ge d$, the density of states at large wavelengths gets too large as the thermodynamic limit is taken. The system cannot tell the infrared modes with $\vert k\vert>0$ apart from the $k=0$ mode, which prevents Bose-Einstein condensation and leads instead to a concentration onto a larger function space than the only $k=0$ mode as $u$ is increased. This is the thermodynamic limit of the one-dimensional problem considered in\ \cite{MD}, in which one typically has $\alpha =2$.

For each regime, we have determined the conditional spectral density giving the average distribution of $u$ among the different Fourier modes of the field. If $\alpha\ge d$, or $\alpha <d$ and $u< u_c$, the conditional spectral density is a smooth function of $k$ given by Eq.\ (\ref{eq3.8}). On the other hand, if $\alpha <d$ and $u> u_c$, the conditional spectral density is given by Eq.\ (\ref{eq3.14}) which clearly shows the condensation into the $k=0$ mode by the presence of the $\delta$-function on its right-hand side.

We have also illustrated the similarities and differences between the condensation properties in the Gaussian-field model and those of the mass transport model in one dimension. While in both cases, there can be both normal and anomalous condensates depending on the parameter regimes, the precise nature of the condensate is different in the two cases. This difference can be traced back to the fact that while the condensation in the mass transport model is homogeneous (in real space), it is heterogeneous (in Fourier space) in the Gaussian-field model.

In conclusion we outline some possible generalizations of this work. As far as scalar fields are concerned, further investigations would involve relaxing some of the assumptions we made about the correlation function. For instance, it would be interesting to investigate how our results are modified when $\tilde{C}(k)$ is not a single bump function of $k$. In the cases where condensation occurs, the condensate is expected to have a much richer functional structure than a single Fourier mode, as it should live in the whole function space spanned by all the Fourier modes for which $\tilde{C}(k)$ is maximum. Such a study would be of interest in e.g. laser-plasma interaction physics in the so-called ``indirect-drive" scheme\ \cite{Lin}. In scalar approximation, the crossing of several spatially incoherent laser beams leads to precisely this kind of Gaussian random field as a model for the laser electric field. An other interesting line of investigation would be to consider vector Gaussian random fields and quadratic forms more general than the simple $L^2$-norm. For instance, in the context of turbulent dynamo\ \cite{Mof}, it would be of great interest to be able to determine the structure(s) of the random flow with a large average helicity over a given scale. Assuming a Gaussian random flow, the question would then be to find out whether there is a concentration onto a smaller flow space when average helicity gets large and whether this concentration turns into a Bose-Einstein condensation when the considered scale gets large. Such studies will be the subject of a future work.
\section*{Acknowledgements}
We thank Alain Comtet for providing valuable insights. Ph. M. also thanks Harvey A. Rose, Pierre Collet, and Joel L. Lebowitz for useful discussions on related subjects.
%
%
%
%
\appendix
\section{Joint probability distribution function of $\bm{{\rm Re}\, \varphi_L(n)}$ and $\bm{ {\rm Im}\, \varphi_L(n)}$}\label{app1}
If $\varphi$ is complex with $\langle\varphi(x)\rangle =\langle\varphi(x)\varphi(y)\rangle =0$ and $\langle\varphi(x)\varphi(y)^\ast\rangle =C(x-y)$, it follows from the definition\ (\ref{cor1.1}) that the $\varphi_L(n)$ are complex Gaussian random variables with
\begin{equation}\label{ap2.1}
\langle\varphi_L(n)\rangle =\langle\varphi_L(n)\varphi_L(m)\rangle =0,
\end{equation}
and
\begin{eqnarray}\label{ap2.2}
\langle\varphi_L(n)\varphi_L(m)^\ast\rangle &=&\frac{1}{V}\int_\Lambda \int_\Lambda
\langle\varphi(x)\varphi(y)^\ast\rangle\, {\rm e}^{-2i\pi (n\cdot x-m\cdot y)}d^dx d^dy \nonumber \\
&=&\frac{1}{V}\int_\Lambda \left\lbrack\int_\Lambda C(x-y)\, {\rm e}^{-2i\pi n\cdot (x-y)}d^dx\right\rbrack
{\rm e}^{2i\pi (m-n)\cdot y} d^dy \nonumber \\
&=&\frac{C_L(n)}{V}\int_\Lambda {\rm e}^{2i\pi (m-n)\cdot y}d^dy =C_L(n)\delta_{n,m},
\end{eqnarray}
where we have used the periodicity of $C(x)$ resulting from the periodic boundary conditions imposed on $\varphi$ and the definition\ (\ref{eq1.1}) of $C_L(n)$. Writing $\varphi_L(n)={\rm Re}\, \varphi_L(n)+i\, {\rm Im}\, \varphi_L(n)$ and $\varphi_L(m)={\rm Re}\, \varphi_L(m)+i\, {\rm Im}\, \varphi_L(m)$ in\ (\ref{ap2.1}) and\ (\ref{ap2.2}), one obtains
\begin{eqnarray}\label{ap2.3}
&&\langle {\rm Re}\, \varphi_L(n)\rangle =\langle {\rm Im}\, \varphi_L(n)\rangle =0, \nonumber \\
&&\langle {\rm Re}\, \varphi_L(n){\rm Im}\, \varphi_L(m)\rangle =0, \\
&&\langle {\rm Re}\, \varphi_L(n){\rm Re}\, \varphi_L(m)\rangle =
\langle {\rm Im}\, \varphi_L(n){\rm Im}\, \varphi_L(m)\rangle =\frac{C_L(n)}{2}\, \delta_{n,m}, \nonumber
\end{eqnarray}
from which it follows that the ${\rm Re}\, \varphi_L(n)$ and the ${\rm Im}\, \varphi_L(n)$ are independent Gaussian random variables with zero mean and variance $C_L(n)/2$. Their joint pdf is thus given by the product measure
\begin{eqnarray}\label{ap2.4}
p\left\lbrack\lbrace {\rm Re}\, \varphi_L(n),\, {\rm Im}\, \varphi_L(n)\rbrace\right\rbrack &=&
\prod_{n\in{\mathbb Z}^d}\mathcal{N}_{\lbrack 0,C_L(n)/2\rbrack}({\rm Re}\, \varphi_L(n))
\times\prod_{n\in{\mathbb Z}^d}\mathcal{N}_{\lbrack 0,C_L(n)/2\rbrack}({\rm Im}\, \varphi_L(n)) \nonumber \\
&=&\prod_{n\in{\mathbb Z}^d}\frac{1}{\pi C_L(n)}\, \exp\left\lbrack -\frac{\vert\varphi_L(n)\vert^2}{C_L(n)}\right\rbrack ,
\end{eqnarray}
where $\mathcal{N}_{\lbrack c_1,c_2\rbrack}(x)$, with $c_2>0$, is the Gaussian measure of mean $c_1$ and variance $c_2$,
\begin{equation}\label{ap2.5}
\mathcal{N}_{\lbrack c_1,c_2\rbrack}(x)=
\frac{1}{\sqrt{2\pi c_2}}\, \exp\left\lbrack -\frac{(x-c_1)^2}{2c_2}\right\rbrack .
\end{equation}

If $\varphi$ is real, its Fourier coefficients are linked by Hermitian symmetry, $\varphi_L(-n)=\varphi_L(n)^\ast$, and $\varphi$ is entirely determined by giving the $\varphi_L(n)$ in only one half of ${\mathbb Z}^d$. Let ${\mathbb Z}_{1/2}^d$ be a given half of ${\mathbb Z}^d$ excluding the point $n=0$. All the statistical properties of $\varphi$ are thus encoded in the joint pdf of $\varphi_L(0)$ (which is necessarily real by Hermitian symmetry), ${\rm Re}\, \varphi_L(n)$, and ${\rm Im}\, \varphi_L(n)$ for $n$ in ${\mathbb Z}_{1/2}^d$. By $\langle\varphi(x)\rangle  =0$, $\langle\varphi(x)\varphi(y)\rangle =C(x-y)$ and the definition\ (\ref{cor1.1}), the $\varphi_L(n)$ for $n$ in $\lbrace 0\rbrace\cup{\mathbb Z}_{1/2}^d$ are complex Gaussian random variables with
\begin{equation}\label{ap2.6}
\langle\varphi_L(n)\rangle =0,
\end{equation}
\begin{eqnarray}\label{ap2.7}
\langle\varphi_L(n)\varphi_L(m)\rangle &=&\frac{1}{V}\int_\Lambda \int_\Lambda
\langle\varphi(x)\varphi(y)\rangle\, {\rm e}^{-2i\pi (n\cdot x+m\cdot y)}d^dx d^dy \nonumber \\
&=&\frac{1}{V}\int_\Lambda \left\lbrack\int_\Lambda C(x-y)\, {\rm e}^{-2i\pi n\cdot (x-y)}d^dx\right\rbrack
{\rm e}^{-2i\pi (m+n)\cdot y} d^dy \nonumber \\
&=&\frac{C_L(n)}{V}\int_\Lambda {\rm e}^{-2i\pi (m+n)\cdot y}d^dy =C_L(n)\delta_{n,-m}=C_L(0)\delta_{n,0}\delta_{m,0},
\end{eqnarray}
($\delta_{n,-m}=\delta_{n,0}\delta_{m,0}$ because both $n$ and $m$ are in $\lbrace 0\rbrace\cup{\mathbb Z}_{1/2}^d$), and
\begin{eqnarray}\label{ap2.8}
\langle\varphi_L(n)\varphi_L(m)^\ast\rangle &=&\frac{1}{V}\int_\Lambda \int_\Lambda
\langle\varphi(x)\varphi(y)\rangle\, {\rm e}^{-2i\pi (n\cdot x-m\cdot y)}d^dx d^dy \nonumber \\
&=&\frac{1}{V}\int_\Lambda \left\lbrack\int_\Lambda C(x-y)\, {\rm e}^{-2i\pi n\cdot (x-y)}d^dx\right\rbrack
{\rm e}^{2i\pi (m-n)\cdot y} d^dy \nonumber \\
&=&\frac{C_L(n)}{V}\int_\Lambda {\rm e}^{2i\pi (m-n)\cdot y}d^dy =C_L(n)\delta_{n,m}.
\end{eqnarray}
Writing $\varphi_L(n)={\rm Re}\, \varphi_L(n)+i\, {\rm Im}\, \varphi_L(n)$ and $\varphi_L(m)={\rm Re}\, \varphi_L(m)+i\, {\rm Im}\, \varphi_L(m)$ in\ (\ref{ap2.6}),\ (\ref{ap2.7}), and\ (\ref{ap2.8}), one obtains
\begin{eqnarray}\label{ap2.9}
&&\langle {\rm Re}\, \varphi_L(n)\rangle =\langle {\rm Im}\, \varphi_L(n)\rangle =0, \nonumber \\
&&\langle {\rm Re}\, \varphi_L(n){\rm Im}\, \varphi_L(m)\rangle =0 \\
&&\langle {\rm Re}\, \varphi_L(n){\rm Re}\, \varphi_L(m)\rangle =\frac{C_L(n)}{2-\delta_{n,0}}\, \delta_{n,m}, \nonumber \\
&&\langle {\rm Im}\, \varphi_L(n){\rm Im}\, \varphi_L(m)\rangle =\frac{C_L(n)}{2}\, (1-\delta_{n,0})\delta_{n,m}, \nonumber
\end{eqnarray}
for $n$ and $m$ in $\lbrace 0\rbrace\cup{\mathbb Z}_{1/2}^d$. It follows that $\varphi_L(0)$, the ${\rm Re}\, \varphi_L(n)$, and the ${\rm Im}\, \varphi_L(n)$ (for $n$ in ${\mathbb Z}_{1/2}^d$) are independent Gaussian random variables with zero mean,  $\langle\varphi_L(0)^2\rangle =C_L(0)$, and $\langle ({\rm Re}\, \varphi_L(n))^2\rangle =\langle ({\rm Im}\, \varphi_L(n))^2\rangle =C_L(n)/2$. Their joint pdf  is therefore given by the product measure
\begin{eqnarray}\label{ap2.10}
&&p\left\lbrack\lbrace \varphi_L(0),\ {\rm Re}\, \varphi_L(n),\, {\rm Im}\, \varphi_L(n)\rbrace\right\rbrack = \nonumber \\
&&\mathcal{N}_{\lbrack 0,C_L(0)\rbrack}(\varphi_L(0))\times
\prod_{n\in{\mathbb Z}_{1/2}^d}\mathcal{N}_{\lbrack 0,C_L(n)/2\rbrack}({\rm Re}\, \varphi_L(n))
\times\prod_{n\in{\mathbb Z}_{1/2}^d}\mathcal{N}_{\lbrack 0,C_L(n)/2\rbrack}({\rm Im}\, \varphi_L(n)) \nonumber \\
&&=\frac{1}{\sqrt{2\pi C_L(0)}}\, \exp\left\lbrack -\frac{\varphi_L(0)^2}{2C_L(0)}\right\rbrack
\times\prod_{n\in{\mathbb Z}_{1/2}^d}\frac{1}{\pi C_L(n)}\, \exp\left\lbrack -\frac{\vert\varphi_L(n)\vert^2}{C_L(n)}\right\rbrack .
\end{eqnarray}
%
%
\section{Calculation of $\bm{R(s,L)}$}\label{app2}
We want to estimate the expression\ (\ref{eq5.11}) of $R(s,L)$ in the limits $L\rightarrow +\infty$, $s\rightarrow 0$, and $L^\alpha s\le O(1)$, with $1/2\le\alpha<1$. Since $s$ tends to zero, on can already simplify\ (\ref{eq5.11}) as
\begin{equation} \label{ap1.1}
R(s,L)=\ln{\left[\varepsilon s \tilde{C}_L(0)\right]} -\frac{2}{\tilde{C}_{L}(0)} \int_{0}^{+\infty} \frac{\lbrack P_1(kL/2\pi)+1/2\rbrack\tilde{C}_L^\prime(k)}{1-\tilde{C}_L(k)/\tilde{C}_L(0)+ \varepsilon s \tilde{C}_{L}(k)}\, dk.
\end{equation}
For any given $\delta >0$ arbitrarily small, we split the $k$-integral on the right-hand side of\ (\ref{ap1.1}) as the sum of an integral from $0$ to $\delta$ plus an integral from $\delta$ to $+\infty$.

The asymptotic behavior of the integral from $\delta$ to $+\infty$ is easily obtained by noting that, in this domain, the derivative of $\ln\lbrack 1-\tilde{C}_L(k)/\tilde{C}_L(0)+ \varepsilon s \tilde{C}_{L}(k)\rbrack$ appearing in the integral is a well behaved function of $k$ when $s\rightarrow 0$ and $L\rightarrow +\infty$. Thus, because of the wild oscillations of $P_1(kL/2\pi)+1/2$ around $1/2$, one gets
\begin{eqnarray}\label{ap1.2}
&&\frac{2}{\tilde{C}_L(0)}
\int_{\delta}^{+\infty}\frac{\lbrack P_1(kL/2\pi)+1/2\rbrack\tilde{C}_L^\prime(k)}{1-\tilde{C}_L(k)/\tilde{C}_L(0)+ \varepsilon s \tilde{C}_{L}(k)}\, dk\sim
\int_{\delta}^{+\infty}\frac{\tilde{C}^\prime(k)/\tilde{C}(0)}{1-\tilde{C}(k)/\tilde{C}(0)}\, dk \nonumber \\
&&=\ln\left\lbrack 1-\frac{\tilde{C}(\delta)}{\tilde{C}(0)}\right\rbrack
\simeq\ln a\delta^\alpha .\ \ \ \ (L\rightarrow +\infty ,\ s\rightarrow 0)
\end{eqnarray}
In Eq.\ (\ref{ap1.2}), as well as in Eqs\ (\ref{ap1.7}) and\ (\ref{ap1.9}) below, the condition $L^\alpha s\le O(1)$ supplementing the limits $L\rightarrow +\infty$ and $s\rightarrow 0$ is always implicitely assumed.

Since $\delta$ is arbitrarily small, we can replace $\tilde{C}_L(k)$ with $\tilde{C}(k)\simeq\tilde{C}(0)(1-ak^\alpha)$ in the integral from $0$ to $\delta$. Choosing $\delta$ such that $\delta L/2\pi$ is an integer, one obtains
\begin{eqnarray}\label{ap1.3}
&& -\frac{2}{\tilde{C}_L(0)}
\int_0^{\delta}\frac{\lbrack P_1(kL/2\pi)+1/2\rbrack\tilde{C}_L^\prime(k)}{1-\tilde{C}_L(k)/\tilde{C}_L(0)+ \varepsilon s \tilde{C}_{L}(k)}\, dk\simeq
2\alpha\int_0^\delta \frac{\lbrack P_1(kL/2\pi)+1/2\rbrack\, ak^{\alpha -1}}{ak^\alpha +\varepsilon s \tilde{C}_L(0)}\, dk \nonumber \\
&&=\frac{\alpha L}{\pi} \int_0^\delta \frac{k^\alpha}{k^\alpha +\eta s}\, dk
-2\alpha\sum_{n=0}^{\delta L/2\pi -1} n\int_{2\pi n/L}^{2\pi (n+1)/L} \frac{k^{\alpha-1}}{k^\alpha +\eta s}\, dk,
\end{eqnarray}
where we have used $P_1(kL/2\pi)+1/2=kL/2\pi -n$ for $2\pi n/L<k<2\pi(n+1)/L$ and written $\eta =\varepsilon\tilde{C}_L(0)/a$. We will now determine the asymptotic behavior of\ (\ref{ap1.3}), first for $1/2<\alpha <1$, then for $\alpha =1/2$.
\subsubsection{$1/2<\alpha<1$}
The behavior of the first term in the second line of\ (\ref{ap1.3}) is given by
\begin{eqnarray}\label{ap1.4}
&&\frac{\alpha L}{\pi} \int_0^\delta \frac{k^\alpha}{k^\alpha +\eta s}\, dk
=\frac{\alpha\delta L}{\pi}\left\lbrack 1-F\left(1,\, \frac{1}{\alpha},\, 1+\frac{1}{\alpha},\, -\frac{\delta^\alpha}{\eta s}\right)\right\rbrack \nonumber \\
&&\sim \frac{\alpha L}{\pi}\left\lbrack\delta -\Gamma\left(1+\frac{1}{\alpha}\right)\Gamma\left(1-\frac{1}{\alpha}\right)(\eta s)^{1/\alpha}
-\frac{\delta^{1-\alpha}}{1-\alpha}\, \eta s\right\rbrack,\ \ \ \ (s\rightarrow 0)
\end{eqnarray}
where $F(a,b,c,z)$ denotes the hypergeometric function ${}_2 F_1(a,b;c;z)$. Performing the integrals in the second term of the second line of\ (\ref{ap1.3}), one gets
\begin{eqnarray}\label{ap1.5}
&&2\alpha\sum_{n=0}^{\delta L/2\pi -1} n\int_{2\pi n/L}^{2\pi (n+1)/L} \frac{k^{\alpha-1}}{k^\alpha +\eta s}\, dk =
2\alpha\sum_{n=0}^{\delta L/2\pi -1} n\left\lbrack\ln\frac{2\pi (n+1)}{L}-\ln\frac{2\pi n}{L}\right\rbrack \nonumber \\
&&+2\sum_{n=0}^{\delta L/2\pi -1} n\left\lbrack\ln\left(1+\frac{L^\alpha\eta s}{\lbrack 2\pi (n+1)\rbrack^\alpha}\right)
-\ln\left(1+\frac{L^\alpha\eta s}{(2\pi n)^\alpha}\right)\right\rbrack \\
&&=\frac{\alpha\delta L}{\pi}\ln\delta -2\alpha\sum_{n=1}^{\delta L/2\pi}\ln\frac{2\pi n}{L}
+\frac{\delta L}{\pi}\ln\left(1+\frac{\eta s}{\delta^\alpha}\right) -2\sum_{n=1}^{\delta L/2\pi}\ln\left(1+\frac{L^\alpha\eta s}{(2\pi n)^\alpha}\right). \nonumber
\end{eqnarray}
It is convenient to introduce the function
\begin{equation}\label{ap1.6}
\phi(t)=\zeta(\alpha)\, t +\sum_{n=1}^{+\infty}\left[\ln\left(1+\frac{t}{n^\alpha}\right) -\frac{t}{n^\alpha} \right ] ,
\end{equation}
in terms of which one has the asymptotics
\begin{eqnarray}\label{ap1.7}
\sum_{n=1}^{\delta L/2\pi}\ln\left(1+\frac{L^\alpha\eta s}{(2\pi n)^\alpha}\right)&\sim& \phi\left(\frac{L^\alpha\eta s}{(2\pi)^\alpha}\right)
- \frac{L^\alpha\eta s}{(2\pi)^\alpha}\, \left(\zeta(\alpha) -\sum_{n=1}^{\delta L/2\pi}\frac{1}{n^\alpha}\right)
\nonumber \\
&\sim& \phi\left(\frac{L^\alpha\eta s}{(2\pi)^\alpha}\right) +\frac{L}{2\pi}\, \frac{\delta^{1-\alpha}}{1-\alpha}\, \eta s\ \ \ \ (L\rightarrow +\infty ,\ s\rightarrow 0),
\end{eqnarray}
where we have used $\sum_{n=1}^{\delta L/2\pi}n^{-\alpha}\sim\zeta(\alpha)+(2\pi/\delta L)^\alpha\lbrack (\delta L/2\pi)/(1-\alpha)+1/2\rbrack$ ($L\rightarrow +\infty$). Putting\ (\ref{ap1.7}), $\ln(1+\eta s/\delta^\alpha)\sim \eta s/\delta^\alpha$ ($s\rightarrow 0$), and
\begin{eqnarray}\label{ap1.8}
&&2\alpha\sum_{n=1}^{\delta L/2\pi}\ln\frac{2\pi n}{L} =\frac{\alpha\delta L}{\pi}\ln\frac{2\pi}{L}
+2\alpha\ln\left(\frac{\delta L}{2\pi}\right)! \nonumber \\
&&\sim \frac{\alpha\delta L}{\pi}\ln\delta +\alpha\ln\delta L -\frac{\alpha\delta L}{\pi},\ \ \ \ (L\rightarrow +\infty)
\end{eqnarray}
in the place of the corresponding terms in the last line of\ (\ref{ap1.5}), one obtains
\begin{eqnarray}\label{ap1.9}
&&2\alpha\sum_{n=0}^{\delta L/2\pi -1} n\int_{2\pi n/L}^{2\pi (n+1)/L} \frac{k^{\alpha-1}}{k^\alpha +\eta s}\, dk \sim
\frac{\alpha L}{\pi}\left(\delta -\frac{\delta^{1-\alpha}}{1-\alpha}\, \eta s\right) \nonumber \\
&&-\left\lbrack\alpha\ln\delta L +2\phi\left(\frac{L^\alpha\eta s}{(2\pi)^\alpha}\right)\right\rbrack .\ \ \ \ (L\rightarrow +\infty ,\ s\rightarrow 0)
\end{eqnarray}
Finally, it follows from\ (\ref{ap1.2}),\ (\ref{ap1.4}), and\ (\ref{ap1.9}) that the asymptotic behavior of\ (\ref{ap1.1}) in the limits $L\rightarrow +\infty$, $s\rightarrow 0$, and $L^\alpha s\le O(1)$ is given by the expression\ (\ref{eq5.13}).
\subsubsection{$\alpha=1/2$}
The behavior of the first term in the second line of\ (\ref{ap1.3}) is now given by
\begin{eqnarray}\label{ap1.10}
&&\frac{L}{2\pi} \int_0^\delta \frac{\sqrt{k}}{\sqrt{k} +\eta s}\, dk
=\frac{\delta L}{2\pi}\left\lbrack 1-F\left(1,\, 2,\, 3,\, -\frac{\delta^{1/2}}{\eta s}\right)\right\rbrack \nonumber \\
&&\sim \frac{L}{2\pi}\left\lbrack\delta -2(\eta s)^2\ln\left(\frac{\eta s}{\delta^{1/2}}\right)
-2\delta^{1/2}\eta s\right\rbrack,\ \ \ \ (s\rightarrow 0)
\end{eqnarray}
and the Eqs. (\ref{ap1.6}) and\ (\ref{ap1.7}) are respectively replaced with
\begin{equation}\label{ap1.11}
\psi(t)=\zeta\left(\frac{1}{2}\right) t -\frac{\gamma t^2}{2}
+\sum_{n=1}^{+\infty}\left\lbrack\ln\left(1+\frac{t}{\sqrt{n}}\right)-\frac{t}{\sqrt{n}}+\frac{t^2}{2n} \right\rbrack ,
\end{equation}
and
\begin{eqnarray}\label{ap1.12}
&&\sum_{n=1}^{\delta L/2\pi}\ln\left(1+\frac{L^{1/2}\eta s}{\sqrt{2\pi n}}\right)\sim \psi\left(\frac{L^{1/2}\eta s}{\sqrt{2\pi}}\right) \nonumber \\
&&- \frac{L^{1/2}\eta s}{\sqrt{2\pi}}\, \left(\zeta(1/2) -\sum_{n=1}^{\delta L/2\pi}\frac{1}{\sqrt{n}}\right)
+\frac{L(\eta s)^2}{4\pi}\, \left(\gamma -\sum_{n=1}^{\delta L/2\pi}\frac{1}{n}\right) \\
&&\sim \psi\left(\frac{L^{1/2}\eta s}{\sqrt{2\pi}}\right) +\frac{\delta^{1/2}L\eta s}{\pi}
-\frac{L(\eta s)^2}{4\pi}\ln\left(\frac{\delta L}{2\pi}\right)\ \ \ \ (L\rightarrow +\infty ,\ s\rightarrow 0), \nonumber
\end{eqnarray}
where we have used $\sum_{n=1}^{\delta L/2\pi}n^{-1/2}\sim\zeta(1/2)+(2\pi/\delta L)^{1/2}(\delta L/\pi+1/2)$ ($L\rightarrow +\infty$) and $\sum_{n=1}^{\delta L/2\pi}n^{-1}\sim\gamma+\ln(\delta L/2\pi)+\pi/(2\delta L)$ ($L\rightarrow +\infty$). As previously, In Eq.\ (\ref{ap1.12}), as well as in Eqs\ (\ref{ap1.13}) below, the condition $L^{1/2} s\le O(1)$ supplementing the limits $L\rightarrow +\infty$ and $s\rightarrow 0$ is implicitely assumed. Putting\ (\ref{ap1.12}),\ (\ref{ap1.8}) with $\alpha =1/2$, and $\ln(1+\eta s/\delta^{1/2})\sim \eta s/\delta^{1/2} -(\eta s)^2/(2\delta)$ ($s\rightarrow 0$) in the place of the corresponding terms in the last line of\ (\ref{ap1.5}) with $\alpha =1/2$, one obtains
\begin{eqnarray}\label{ap1.13}
&&\sum_{n=0}^{\delta L/2\pi -1} n\int_{2\pi n/L}^{2\pi (n+1)/L} \frac{1}{\sqrt{k} +\eta s}\, \frac{dk}{\sqrt{k}} \sim
-\left\lbrack\frac{1}{2}\ln\delta L +2\psi\left(\frac{L^{1/2}\eta s}{\sqrt{2\pi}}\right)\right\rbrack \nonumber \\
&&+\frac{L}{2\pi}\left\lbrack\delta -2\delta^{1/2} \eta s -(\eta s)^2\right\rbrack
+\frac{L}{2\pi}\, (\eta s)^2\, \ln\left(\frac{\delta L}{2\pi}\right).\ \ \ \ (L\rightarrow +\infty ,\ s\rightarrow 0)
\end{eqnarray}
Finally, using \ (\ref{ap1.2}),\ (\ref{ap1.10}),\ (\ref{ap1.13}), and $K=\varepsilon\tilde{C}(0)^2/(\pi a^2)=\eta^2/(\varepsilon\pi)$ [see Eq.\ (\ref{eq2.11})] in \ (\ref{ap1.1}), one finds that for $\alpha =1/2$ the asymptotic behavior of $R(s,L)$ in the limits $L\rightarrow +\infty$, $s\rightarrow 0$, and $L^{1/2} s\le O(1)$ is given by the expression\ (\ref{eq5.16}).
%
%

%
%
\end{document}